\documentclass[a4paper,fleqn]{article}
\usepackage[T1]{fontenc}
\usepackage{stix}
\usepackage{microtype}

\usepackage[
  paperwidth=192mm,
  paperheight=262mm,
  hmargin=13.7mm,
  vmargin=19mm
]{geometry}
\usepackage[authoryear,longnamesfirst]{natbib}
\usepackage{bbm}
\usepackage{amsmath,amsfonts,amssymb,amsxtra,amsthm,amstext,bbm}
\usepackage{graphicx}
\usepackage{url}

\newcommand{\card}[1]{|#1|}

\newcommand{\onevect}[0]{\mathbf{1}}
\newcommand{\variance}[0]{\mathrm{var}}

\newcommand{\mrcaAgeRVU}[1][\treeRV]{\bar{A}^{\scalebox{.3}[.5]{$\mathrm{MRCA}$}}_{#1}}
\newcommand{\mrcaAgeVa}[1][\tree]{A^{\scalebox{.3}[.5]{$\mathrm{MRCA}$}}_{#1}}
\newcommand{\effRec}[1][\tree]{N_{#1}}
\newcommand{\pSurvival}[0]{p}

\newcommand{\mexp}[1]{e^{#1}}
\newcommand{\mder}[0]{d}
\newcommand{\ftime}[1][\ttot]{\phi_{#1}}

\newcommand{\rebirth}[1][\ttot]{\tilde{\birth}_{#1}}
\newcommand{\tree}[0]{{\mathcal{T}}}
\newcommand{\treeRV}[0]{\dot{\mathcal{T}}}
\newcommand{\phyvar}[1][\tree]{C_{#1}}

\newcommand{\rtree}[1][\tree]{r}

\newcommand{\tips}[1][\tree]{\mathrm{L}_{#1}}

\newcommand{\ttot}[0]{t}
\newcommand{\tfir}[0]{s}

\newcommand{\tdiv}[0]{t_{\mathrm{d}}}
\newcommand{\tspec}[0]{t_{\mathrm{s}}}
\newcommand{\tspecBef}[0]{t^{-}_{\mathrm{s}}}
\newcommand{\tspecAft}[0]{t^{+}_{\mathrm{s}}}
\newcommand{\texti}[0]{t_{\mathrm{e}}}
\newcommand{\textiBef}[0]{t^{-}_{\mathrm{e}}}
\newcommand{\textiAft}[0]{t^{+}_{\mathrm{e}}}
\newcommand{\trait}[1][\tree(\ttot)]{X_{#1}}
\newcommand{\meanTrait}[1][\tree(\ttot)]{\bar{X}_{#1}}
\newcommand{\transpose}[1]{#1'}
\newcommand{\matProj}[1][\tree(\ttot)]{P_{#1}}
\newcommand{\ident}[0]{I}

\newcommand{\birth}[0]{\lambda}
\newcommand{\death}[0]{\mu}

\newcommand{\prob}[0]{\mathbbm{P}}
\newcommand{\esp}[0]{\mathbbm{E}}
\newcommand{\mavar}[1]{S^{2}_{#1}}

\newcommand{\distri}[1]{F_{#1}}
\newcommand{\densi}[1]{f_{#1}}
\newtheorem{proposition}{Proposition}
\newtheorem{corollary}{Corollary}

\title{Phylogenetic dynamics of MRCA ages and empirical moments of a Brownian trait}
\author{Gilles Didier\\
\small Institut Montpelliérain Alexander Grothendiek\\
\small Université de Montpellier, Case courrier 051\\
\small Place Eugène Bataillon,  34090 Montpellier, France\\
\small \texttt{gilles.didier@umontpellier.fr}
}

\begin{document}
\maketitle
\begin{abstract}
We study the temporal dynamics of the first two empirical moments of Brownian
traits on phylogenetic trees. For a fixed tree, we characterize the
distributions of their empirical mean and empirical variance across all lineages
extant at any given time. In particular, we show that the variance of the empirical mean and the expected empirical variance are piecewise linear between diversification events.

For lineage-homogeneous random trees, both the variance of
the empirical mean and the expected empirical variance can be expressed in
terms of the expected age of the most recent common ancestor (MRCA) of a uniformly
sampled pair of extant lineages. In this representation, the expected MRCA age enters
the two quantities with opposite signs, pointing to a structural opposition between
the variance of the empirical mean and the expected empirical variance.

For generalized birth-death processes with time-dependent speciation and
extinction rates, we derive an explicit formula for the distribution of the MRCA age of a uniformly
sampled pair of extant lineages. This yields integral expressions, at any time,
for both the variance of the empirical mean and the expected empirical variance.
In the constant-rate birth-death case, we further obtain closed-form expressions
for the expected empirical variance and describe its asymptotic behavior in the
supercritical, critical and subcritical regimes.
\end{abstract}


\section{Introduction}

A stochastic model of quantitative trait evolution specifies how
trait values change along lineages. At any given time, this lineage-level process
also induces a collection of trait values across the lineages then present.
Such a collection, like any set of observations, can often be more readily
interpreted through summary quantities that describe its distribution than
through the raw list of individual values. The empirical mean and the
empirical variance are standard such quantities: they describe, respectively,
the average position of the clade in trait space and the spread of trait values
around this average. In this sense, the empirical variance provides one natural
measure of trait disparity among several others used in macroevolutionary
studies~\citep{Wills1994}.

\begin{figure}
\centerline{\includegraphics[width=0.66\textwidth]{"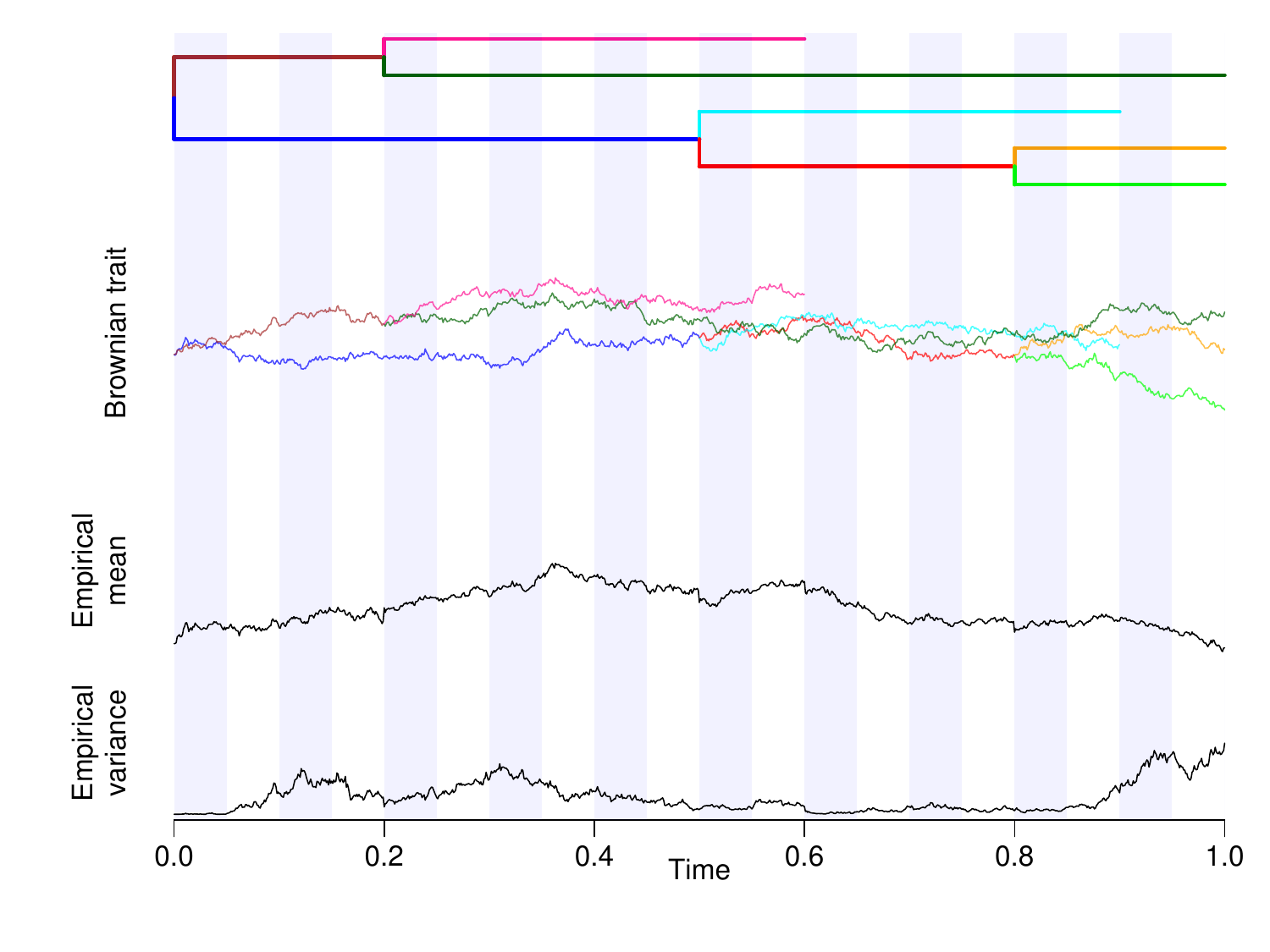"}}
\caption{A phylogenetic tree (top). Simulated evolution of a quantitative trait along this tree under a Brownian motion model (center). Empirical variance of trait values computed across all lineages present at each time point (bottom).}\label{fig:simulation}
\end{figure}

If trait values are assumed to follow a stochastic process along lineages,
the empirical mean and the empirical variance computed across the lineages
present at each time also evolve as derived stochastic processes (Fig.~\ref{fig:simulation}). 
A key point is that these derived processes are not determined by the
lineage-level dynamics alone. They also depend on the phylogeny, both through
the number of lineages present at a given time and through the dependence
induced by shared ancestry among their trait values. This leads to two related but distinct questions: how these
empirical moments behave on a given fixed tree, and how their behavior is
affected when the tree itself is generated by a stochastic diversification
process. For Brownian traits, the second question will turn out to be closely related to
the dynamics of the mean most recent common ancestor (MRCA) age of pairs of
extant lineages.

Several related aspects of these questions have been studied previously, notably in the Brownian case.
For a fixed phylogeny, the covariance structure induced by Brownian motion is
determined by the phylogenetic variance-covariance matrix, and quantities such
as the empirical mean, the empirical variance, or trait disparity can be
expressed in terms of this matrix, whose entries are the MRCA ages of the considered lineages. This connection has been used
in phylogenetic comparative methods and in models testing for variation in
rates of continuous trait evolution
\citep{Felsenstein1985,Grafen1989,OMeara2006}. Other works have considered
random trees and have related the
expected empirical variance to MRCA ages mainly under Yule models
\citep{Sagitov2012,Crawford2013,Bartoszek2015,Mulder2015}. 
Among these works, \citet{Sagitov2012} and
\citet{Mulder2015} are particularly relevant to the
present study, although from different perspectives. \citet{Sagitov2012}
derived expressions for the variance of the empirical mean and the expected
empirical variance of Brownian traits on constant-rate birth-death
trees by conditioning on the number of extant species. Their approach is therefore closely related to the present one, but it concerns
summary statistics conditional on a fixed number of extant species rather than
the temporal dynamics of the corresponding empirical processes. \citet{Mulder2015} studied Markov models of character evolution, including
Brownian motion, on Yule trees from a more dynamic viewpoint, but still
conditionally on the number of lineages and without treating the empirical mean
and empirical variance themselves as stochastic processes.
Disparity-through-time approaches are also related in spirit, since they study
how a summary of morphological variation changes through the history of a clade
\citep{Foote1993,Foote1997,Harmon2003}. However, the disparity considered in
this literature is generally a broader and more complex quantity, based on the
overall morphology of lineages, rather than the empirical variance of a single
trait.

The aim of the present work is to characterize the distributional dynamics of
the empirical mean and empirical variance of Brownian traits. We first consider a
fixed phylogenetic tree and describe, at every time, the distribution of the
empirical mean and empirical variance of Brownian trait values across extant
lineages. The empirical mean is Gaussian, with mean equal to the trait value at the root and variance determined by
MRCA ages and the Brownian variance parameter, whereas the empirical variance is a central quadratic form in a
Gaussian vector. These characterizations provide access to expectations, distributions
and quantiles for both empirical moments. In particular, the variance of the empirical
mean and the expected empirical variance are piecewise linear between
diversification events and may undergo jumps at these events.

We then turn to random trees. For lineage-homogeneous tree processes, exchangeability of extant lineages
implies that the effect of diversification on these quantities is mediated by
the MRCA age of a uniformly sampled pair of distinct extant tips. This yields general expressions for the variance
of the empirical mean and the expected empirical variance.
These expressions reveal a structural contrast between the two quantities:
at fixed number of extant lineages, the expected MRCA age of a uniformly
sampled pair contributes positively to the variance of the empirical mean and
negatively to the expected empirical variance. This suggests that diversification scenarios
favoring displacement of the empirical mean tend to be associated with lower
expected trait disparity, and conversely.

Finally, we specialize these results to generalized birth-death processes with time-dependent speciation and extinction rates. Using the reconstructed
process conditioned on survival up to the observation time, and applying a
deterministic time change, the distribution of the MRCA age of a uniformly
sampled pair can be
derived from a unit-rate Yule process. Combining this distribution with the
previous moment identities gives one-dimensional integral expressions for the
variance of the empirical mean and for the expected empirical variance.
In the constant-rate case, the expressions for the expected empirical variance simplify further, yielding closed forms, involving the dilogarithm function, and explicit asymptotic regimes.

The paper is organized as follows. Section~\ref{sec:fixed} treats fixed trees
and characterizes the empirical mean and empirical variance across extant
lineages, including their behavior between and at diversification events.
Section~\ref{sec:linHom} relates these quantities to MRCA ages for
lineage-homogeneous random trees. Section~\ref{sec:birth_death} applies this
framework to generalized birth-death processes and analyzes the constant-rate
case. The Discussion returns to the biological interpretation of these results,
with particular emphasis on the opposition between displacement of the
empirical mean and expected trait disparity. An R package implementing the results presented here is available at
\url{https://github.com/gilles-didier/PhyloTraitDynamics}, together with the
R script used to generate the figures.

\section{Fixed trees}\label{sec:fixed}

To study the dynamics of morphological disparity under Brownian motion on a
fixed phylogenetic tree, we begin with the elementary case of $n$ independent
lineages, corresponding to a star tree with $n$ tips. This case provides a
useful reference point for the general phylogenetic setting.

\subsection{Independent Brownian motions - Star-tree case}\label{sec:ind}
Assume that, for some positive integer $n$, we have
$\trait[\ttot]^{(i)} = a+\sigma B_i(\ttot)$ for all $i=1,\ldots,n$ and all
$\ttot\geq 0$, where $\sigma>0$ and $B_1,\ldots,B_n$ are independent standard Brownian
motions.

It follows that $(\trait[\ttot]^{(i)})_{i=1,\ldots, n}$ are independent Gaussian random variables with mean $a$ and variance $\sigma^{2}\ttot$, which directly implies that the empirical mean at time $\ttot$, defined as 
\[\meanTrait[\ttot]=\frac{\sum_{i=1}^{n}\trait[\ttot]^{(i)}}{n},\]
is a Gaussian random variable with mean $a$ and variance $\frac{\sigma^{2}\ttot}{n}$.

For all $n\geq 2$, the empirical variance at time $\ttot$ of the joint process $(\trait[\ttot]^{(i)})_{i=1,\ldots, n}$ is defined as 
\[
\mavar{n}(\ttot) = \frac{1}{n-1}\sum_{i=1}^{n}(\trait[\ttot]^{(i)}-\meanTrait[\ttot])^{2}.
\] 
By convention, if $n<2$, we set $\mavar{n}(\ttot) = 0$.

By \citet[Theorem 5.3.1, p.~178]{Casella2002}, $\frac{(n-1)\mavar{n}(\ttot)}{\sigma^{2}\ttot}$ follows a $\chi^{2}$ distribution with $n-1$ degrees of freedom. 

The expectation of $\mavar{n}(\ttot)$ is \[\esp(\mavar{n}(\ttot)) = \frac{\sigma^{2}\ttot\esp(\chi^{2}_{n-1})}{n-1} = \sigma^{2}\ttot,\] which increases linearly with time. 

Its variance is \[\variance(\mavar{n}(\ttot)) = \frac{\sigma^{4}\ttot^{2}\variance(\chi^{2}_{n-1})}{(n-1)^{2}}=\frac{2\sigma^{4}\ttot^{2}}{n-1}.\]

\subsection{Phylogenetic case}\label{sec:fixed:phylo}
\subsubsection{Definitions and notations}\label{sec:fixed:phylo:background}
Let $\tree$ be a rooted phylogenetic tree with branch lengths, and let
$\tips$ denote the set of its tips. Unless specified otherwise, the tips of
$\tree$ are labelled by $1,\ldots,\card{\tips}$, where $\card{A}$ denotes
the cardinality of any finite set $A$. Unless otherwise specified, the tree $\tree$ is not assumed to be ultrametric.

For any node $k$, including tips, the age
of $k$ is obtained by summing the branch lengths along the path from the
root to $k$.

For any two nodes $k$ and $\ell$ of $\tree$, the most recent
common ancestor (MRCA) of $k$ and $\ell$ is the unique common ancestor of
$k$ and $\ell$ that is a descendant of all their common ancestors, with the
usual convention that the MRCA of a node with itself is the node itself. We
denote by $\mrcaAgeVa(k,\ell)$ the MRCA age of $k$ and $\ell$, that is,
the age of their MRCA. In particular, for all nodes $k$,
$\mrcaAgeVa(k,k)$ is the age of $k$.

We write $\phyvar$ for the phylogenetic variance-covariance matrix associated with $\tree$,
indexed by the tips of $\tree$~\citep{Felsenstein1985, Grafen1989}. Its entries are $\phyvar(i,j)=\mrcaAgeVa(i, j)$ for any pair $(i,j)$ of tips.

For any time $\ttot$ within the temporal span of $\tree$, let $\tree(\ttot)$ be the
tree obtained by truncating $\tree$ at time $\ttot$ and removing all lineages that
became extinct before that time. This is the reconstructed tree at time $\ttot$
in the sense of \citet{Nee1994}. Note that $\tree(\ttot)$ is ultrametric with height $\ttot$ by construction. 

Let $\trait[\tree(\ttot)] = (\trait[\tree(\ttot)]^{(i)})_{i\in \tips[\tree(\ttot)]}$ be the trait value vector of the lineages extant at $\ttot$, 
which are the tips of $\tree(\ttot)$. Under a Brownian motion with variance parameter $\sigma^{2}$ and assuming an initial value $a$, 
the random trait-vector $\trait[\tree(\ttot)]$ follows a multivariate normal distribution with mean $a\onevect$ and covariance matrix $\sigma^{2}\phyvar[\tree(\ttot)]$, 
where $\onevect$ denotes the vector of ones of appropriate dimension~\citep{Bastide2024}.

\subsubsection{Empirical mean}\label{sec:fixed:phylo:mean}
If $\card{\tips[\tree(\ttot)]}\geq 1$, the empirical mean at time $\ttot$ (Figure~\ref{fig:simulation}, middle panel) is a linear combination of the Gaussian random vector of trait values $\trait[\tree(\ttot)]$,
namely,
\[
\meanTrait[\tree(\ttot)] =\frac{\sum_{i=1}^{\card{\tips[\tree(\ttot)]}}\trait[\tree(\ttot)]^{(i)}}{\card{\tips[\tree(\ttot)]}}
=
\frac{\transpose{\onevect}\trait[\tree(\ttot)] }{\card{\tips[\tree(\ttot)]}}.
\]
It is therefore a Gaussian
random variable with mean $a$ and variance
\[
\variance(\meanTrait[\tree(\ttot)])
=
\frac{\sigma^2}{\card{\tips[\tree(\ttot)]}^2}
\transpose{\onevect}\phyvar[\tree(\ttot)]\onevect
=
\frac{\sigma^2}{\card{\tips[\tree(\ttot)]}^2}
\sum_{i,j\in\tips[\tree(\ttot)]}\mrcaAgeVa[\tree(\ttot)](i,j).
\]
Since $\tree(\ttot)$ is ultrametric with height $\ttot$, the diagonal terms satisfy
$\mrcaAgeVa[\tree(\ttot)](i,i)=\ttot$ for all tips $i$. It follows that
\[
\variance(\meanTrait[\tree(\ttot)])
=
\frac{\sigma^2}{\card{\tips[\tree(\ttot)]}^2}
\left(
\card{\tips[\tree(\ttot)]}\ttot
+
\sum_{i,j\in\tips[\tree(\ttot)], i\neq j}\mrcaAgeVa[\tree(\ttot)](i,j)
\right).
\]

\begin{figure}
\centerline{\includegraphics[width=0.66\textwidth]{"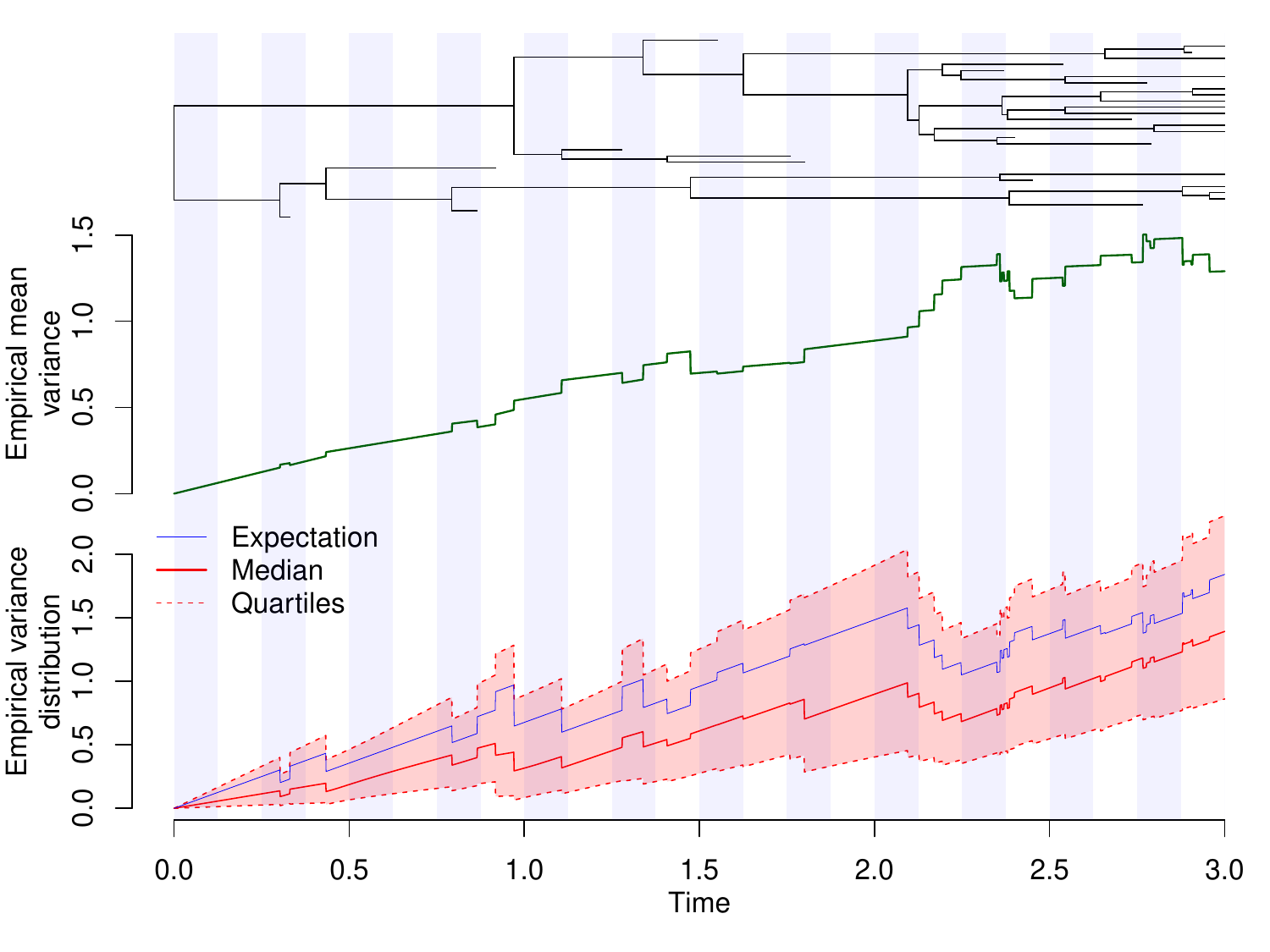"}}
\caption{Top: a non-ultrametric phylogenetic tree. Middle: variance of the empirical mean of trait values evolving along this tree under a standard  Brownian motion. Bottom: expectation, median and quartiles of the empirical variance of these trait values.}
\label{fig:simTreeDist}
\end{figure}

Since $\meanTrait[\tree(\ttot)]$ is Gaussian and its
mean is always equal to the initial value of the Brownian motion, its distribution at
time $\ttot$ is fully characterized by its variance. The middle panel of
Figure~\ref{fig:simTreeDist} displays the dynamics of the variance of the empirical mean of trait
values evolving along a phylogenetic tree under a standard Brownian motion.
The variance appears to increase linearly between
speciation and extinction times, with possible jumps at these times. This
observation is formalized by the following proposition.

\begin{proposition}\label{prop:piecewiseMean}
Let $\tree$ be a phylogenetic tree, and assume that the trait evolves along
$\tree$ according to a Brownian motion with variance parameter $\sigma^{2}$.
Then the variance of the empirical mean $\meanTrait[\tree(\ttot)]$ is piecewise
linear between diversification events, with slope $\sigma^{2}/n$ on any time interval
during which the number of extant lineages is equal to $n\geq 1$.
\end{proposition}

It follows that the variance of the empirical
mean is essentially driven by the diversification history of the tree.

\subsubsection{Empirical variance}\label{sec:fixed:phylo:var}

Under the same assumptions, if $\card{\tips[\tree(\ttot)]}\geq 2$, the empirical
variance across extant lineages at time $\ttot$ is defined by
\[
\mavar{\tree(\ttot)}
=
\frac{1}{\card{\tips[\tree(\ttot)]}-1}
\sum_{i\in \tips[\tree(\ttot)]}
(\trait[\tree(\ttot)]^{(i)}-\meanTrait[\tree(\ttot)])^{2}.
\]
If fewer than two lineages are extant, we set $\mavar{\tree(\ttot)}=0$. This quantity is illustrated in Figure~\ref{fig:simulation}, bottom panel.

Equivalently,
\[
(\card{\tips[\tree(\ttot)]}-1)\mavar{\tree(\ttot)}
=
\transpose{\trait[\tree(\ttot)]}\matProj[\tree(\ttot)] \trait[\tree(\ttot)],
\]
where
\[
\matProj[\tree(\ttot)]
=
\ident
-
\frac{\onevect\transpose{\onevect}}{\card{\tips[\tree(\ttot)]}}.
\]
Thus $(\card{\tips[\tree(\ttot)]}-1)\mavar{\tree(\ttot)}$ is a quadratic form in the
Gaussian vector $\trait[\tree(\ttot)]$.

Since $\esp(\trait[\tree(\ttot)])=a\onevect$ 
and since $\matProj[\tree(\ttot)]\onevect=0$, we have 
\[
(\card{\tips[\tree(\ttot)]}-1)\mavar{\tree(\ttot)}
=
(\transpose{\trait[\tree(\ttot)]}-a\transpose{\onevect})\matProj[\tree(\ttot)] (\trait[\tree(\ttot)]-a\onevect).
\]
This is a central quadratic form, that is, a quadratic form in a centered random vector. 
Its distribution is therefore given by
\[
(\card{\tips[\tree(\ttot)]}-1)\mavar{\tree(\ttot)}
\stackrel{d}{=}
\sum_{i\in \tips[\tree(\ttot)]} \lambda_i Z_i^2,
\]
where the $Z_i$ are independent standard normal random variables and the
$\lambda_i$ are the eigenvalues of
\[
\sigma^{2}
\phyvar[\tree(\ttot)]^{\frac{1}{2}}
\matProj[\tree(\ttot)]
\phyvar[\tree(\ttot)]^{\frac{1}{2}},
\]
\citep[see][p.~90]{Mathai1992}. Equivalently,
$(\card{\tips[\tree(\ttot)]}-1)\mavar{\tree(\ttot)}$ is a weighted sum of independent
$\chi^{2}$ random variables with one degree of freedom.

Several methods have been developed to compute the distribution of such sums
\citep{Imhof1961,Davies1980}, and practical implementations are available
\citep{Duchesne2010}. As a consequence, the distribution of the empirical
variance, and hence quantities such as its expectation, variance and quantiles,
can be computed numerically at any time $\ttot$ within the temporal span of the phylogenetic
tree.

The bottom panel of Figure~\ref{fig:simTreeDist} illustrates the time evolution of the distribution of the empirical variance of a
Brownian trait across extant lineages along the tree at the top of the figure.
The expectation appears piecewise linear; this is proved below.

\begin{proposition}\label{prop:piecewiseVar}
Let $\tree$ be a phylogenetic tree, and assume that the trait evolves along
$\tree$ according to a Brownian motion with variance parameter $\sigma^{2}$.
Then the expectation of the empirical variance $\mavar{\tree(\ttot)}$ is piecewise
linear between diversification events, with slope $\sigma^{2}$ on any time interval during which at
least two lineages are extant.
\end{proposition}

In other words, the expected disparity dynamics consists of linear increases
of slope $\sigma^{2}$, punctuated by jumps at diversification events.

The median and quartiles appear to follow a similar qualitative behavior in
this example: on each interval between diversification events, their
trajectories appear linear, while speciation and extinction times are associated with
jumps and, sometimes, visible changes in slope. The full distribution of the
empirical variance depends strongly on tree topology and branch lengths. In particular, this
makes its analysis for random trees difficult.

In \citet{Sagitov2012} and \citet{Crawford2013}, the authors derive an explicit
expression for the expected empirical variance of the tip values on an
ultrametric tree as a function of its phylogenetic covariance matrix, or,
equivalently, of its MRCA ages. This formula will be useful below.

\begin{proposition}[\citet{Sagitov2012,Crawford2013}]
\label{prop:exp_cov}
Let $\tree$ be an ultrametric phylogenetic tree with height $\ttot$, and let
$\mavar{\tree}$ be the empirical variance at its tips of a trait evolving along $\tree$ according to a
Brownian motion with variance parameter $\sigma^{2}$. Then
\[
\esp(\mavar{\tree})
=
\left\{
\begin{array}{ll}
\sigma^{2}\left(
\ttot-\frac{2\sum_{1\leq i<j\leq\card{\tips[\tree]}}\mrcaAgeVa(i,j)}
{\card{\tips[\tree]}(\card{\tips[\tree]}-1)}
\right)
& \mbox{if $\card{\tips[\tree]}\geq 2$,}\\
0
& \mbox{otherwise.}
\end{array}
\right.
\]
\end{proposition}

The formulas in \citet{Sagitov2012} and \citet{Crawford2013} differ slightly but
are algebraically equivalent, except that the latter normalizes the empirical
variance by the number of tips rather than by the number of tips minus one.

\section{Lineage-homogeneous random trees}\label{sec:linHom}
A random tree process, and by extension the resulting random trees, is said to be \emph{lineage-homogeneous} if a speciation or an extinction event taking place at a given time is attached to any lineage present at that time with uniform probability.

In particular, lineage homogeneity constrains the average behavior of the jumps in the variance of the empirical mean and in the expected empirical variance, observed in Figure~\ref{fig:simTreeDist} and pointed out in Propositions~\ref{prop:piecewiseMean} and~\ref{prop:piecewiseVar}. Namely, the expectation of the jumps in the variance of the empirical mean at speciation and extinction events is nonnegative when conditioning on having at least one and at least two extant lineages at the event time, respectively. By contrast, the expected jump in the expected empirical variance at speciation and extinction events is nonpositive and zero, respectively, when conditioning on having at least two and at least three extant lineages at the event time (Appendix~\ref{app:jumps}).

Below we shall investigate how both the variance of the
empirical mean and the expected empirical variance of a trait evolving under
Brownian motion on lineage-homogeneous random trees are related to the expected MRCA age
of a uniformly chosen pair of tips.

\subsection{Exchangeability of MRCA ages}\label{sec:linhom:exchang}
The lineage homogeneity of the random tree $\treeRV$ implies that, for every
$n\geq 1$, conditionally on
$\card{\tips[\treeRV(\ttot)]}=n$,
the law of the (truncated) tree $\treeRV(\ttot)$ is invariant under permutations
of the $n$ tip labels. A direct consequence of this fact is that, conditionally
on $\card{\tips[\treeRV(\ttot)]}=n$, for all $n\geq 2$ and all
$1\leq i\neq j\leq n$, the random variables
$\mrcaAgeVa[\treeRV(\ttot)](i,j)$ are identically distributed.

Let us define $\mrcaAgeRVU[\treeRV(\ttot)]$ as the MRCA age of a pair 
uniformly chosen among all unordered pairs of distinct tips of $\treeRV(\ttot)$.
Since the definition of the random variable $\mrcaAgeRVU[\treeRV(\ttot)]$ makes
sense only if $\treeRV(\ttot)$ has more than one tip, we shall consider its law,
expectation, and related quantities only in conjunction with, or conditionally on,
events implying $\card{\tips[\treeRV(\ttot)]}\geq 2$. Since the MRCA ages are
identically distributed conditionally on the number of tips, for all $n\geq 2$
and all $1\leq i\neq j\leq n$,
\[
\esp\left(
\mrcaAgeVa[\treeRV(\ttot)](i,j)
\mid
\card{\tips[\treeRV(\ttot)]}= n
\right)
=
\esp(\mrcaAgeRVU[\treeRV(\ttot)]\mid
\card{\tips[\treeRV(\ttot)]}= n).
\]
Note that, since $\treeRV(\ttot)$ is ultrametric of height $\ttot$, we also have
$
\esp(\mrcaAgeRVU[\treeRV(\ttot)]\mid
\card{\tips[\treeRV(\ttot)]}= n)\leq \ttot
$
for all $n\geq 2$.

\subsection{Variance of the empirical mean and MRCA ages}

Let $\treeRV(\ttot)$ be a lineage homogeneous random ultrametric tree of height $\ttot$. Conditionally on
$\card{\tips[\treeRV(\ttot)]}=n$, its tips are labelled by $1,\ldots,n$, and the
law of $\treeRV(\ttot)$ is invariant under permutations of these
labels. We consider a trait evolving under Brownian motion with variance
parameter $\sigma^{2}$ along $\treeRV(\ttot)$.
 In Section~\ref{sec:fixed}, we
considered the variance of the empirical mean of the tip trait values for a
fixed tree: the only source of randomness was then the trait evolution.
Here, the variance is taken with respect to both sources of randomness: the
random tree, conditionally on its number of tips, and the Brownian motion on
this tree. By the law of total variance, conditionnally on $\card{\tips[\treeRV(\ttot)]}=n$, for all $n\geq 1$, we get
\[
\variance(\meanTrait[\treeRV(\ttot)]\mid \card{\tips[\treeRV(\ttot)]}=n)
=
\esp\left(
\variance(\meanTrait[\treeRV(\ttot)]\mid \treeRV(\ttot))
\mid
\card{\tips[\treeRV(\ttot)]}=n
\right)
+
\variance\left(
\esp(\meanTrait[\treeRV(\ttot)]\mid \treeRV(\ttot))
\mid
\card{\tips[\treeRV(\ttot)]}=n
\right).
\]
Since we always have $\esp(\meanTrait[\treeRV(\ttot)]\mid \treeRV(\ttot))=a$, the second term is zero.
Therefore
\begin{equation}\label{eq:varMean}
\variance(\meanTrait[\treeRV(\ttot)]\mid \card{\tips[\treeRV(\ttot)]}=n)
=
\esp\left(
\variance(\meanTrait[\treeRV(\ttot)]\mid \treeRV)
\mid
\card{\tips[\treeRV(\ttot)]}=n
\right).
\end{equation}

For a fixed realization of $\treeRV(\ttot)$ with $n\geq 1$ tips, we have
\[
\variance(\meanTrait[\treeRV(\ttot)]\mid \treeRV(\ttot))
=
\frac{\sigma^2}{n^2}
\left(
n\ttot
+
\sum_{1\leq i\neq j\leq n}
\mrcaAgeVa[\treeRV(\ttot)](i,j)
\right).
\]
Note that, if $n=1$, the sum over MRCA ages is empty and is therefore equal to zero. We have
\begin{equation*}
\variance(\meanTrait[\treeRV(\ttot)]\mid \card{\tips[\treeRV(\ttot)]}=n)
=
\frac{\sigma^2}{n^2}
\left(
n\ttot
+
\sum_{1\leq i\neq j\leq n}
\esp\left(
\mrcaAgeVa[\treeRV(\ttot)](i,j)
\mid
\card{\tips[\treeRV(\ttot)]}=n
\right)
\right).
\end{equation*}

Using the property and the notation introduced in Section~\ref{sec:linhom:exchang}, we obtain, for $n\geq 1$,
\begin{equation}\label{eq:meanVarianceCond}
\variance(\meanTrait[\treeRV(\ttot)]\mid \card{\tips[\treeRV(\ttot)]}=n)
=\left\{\begin{array}{ll}
\frac{\sigma^2}{n}\left(\ttot+(n-1)\esp(\mrcaAgeRVU[\treeRV(\ttot)]\mid \card{\tips[\treeRV(\ttot)]}=n)\right) & \mbox{if $n\geq 2$,}\\
\sigma^{2}\ttot & \mbox{if $n=1$.}\end{array}\right.
\end{equation}

In words, the variance of the empirical mean can be expressed in terms of the
expected MRCA age of a uniformly chosen pair of distinct tips.
\subsection{Expected empirical variance and MRCA ages}

This section builds on ideas developed in
\citet{Sagitov2012,Mulder2015,Crawford2013}, which relate the expected
empirical variance to MRCA ages.

We use the same setting as in the previous subsection, i.e., $\treeRV(\ttot)$ is still a
lineage-homogeneous random ultrametric tree of height $\ttot$, with tips labelled by $1,\ldots,n$
conditionally on $\card{\tips[\treeRV(\ttot)]}=n$.

Applying Proposition~\ref{prop:exp_cov} conditionally on the number of tips
yields
\begin{equation*}
\esp(\mavar{\treeRV(\ttot)}{\mid} \card{\tips[\treeRV(\ttot)]}{=} n)
=
\left\{
\begin{array}{ll}
\sigma^{2}\ttot
-
\frac{2\sigma^{2}}{n(n-1)}
\sum_{1\leq i<j\leq n}
\esp(\mrcaAgeVa[\treeRV(\ttot)](i,j){\mid} \card{\tips[\treeRV(\ttot)]}{=} n)
& \mbox{if $n\geq 2$,}\\
0 & \mbox{otherwise.}
\end{array}
\right.
\end{equation*}

Using the exchangeability property and the notation introduced above, we get
\begin{equation}\label{eq:varExpectCond}
\esp(\mavar{\treeRV(\ttot)}{\mid} \card{\tips[\treeRV(\ttot)]}{=} n)
=
\left\{
\begin{array}{ll}
\sigma^{2}\left(\ttot-\esp(\mrcaAgeRVU[\treeRV(\ttot)]{\mid} \card{\tips[\treeRV(\ttot)]}{=} n)\right)
& \mbox{if $n\geq 2$,}\\
0 & \mbox{otherwise.}
\end{array}
\right.
\end{equation}
This identity gives a meaningful interpretation of expected disparity:
conditionally on the number of extant tips, it is the expected Brownian
variance accumulated since the MRCA of a uniformly chosen pair of distinct
extant lineages, under the random tree distribution.

The expectation of the empirical variance at time $\ttot$, without conditioning on
the number of tips, is then
\begin{align}
\esp\left(\mavar{\treeRV(\ttot)}\right)
&=
\sum_{n=2}^{\infty}
\esp(\mavar{\treeRV(\ttot)}{\mid} \card{\tips[\treeRV(\ttot)]}{=} n)
\prob(\card{\tips[\treeRV(\ttot)]}{=} n)
\notag\\
&=
\sigma^{2}\left(
\ttot\prob(\card{\tips[\treeRV(\ttot)]}{\geq} 2)
-
\sum_{n=2}^{\infty}
\esp(\mrcaAgeRVU[\treeRV(\ttot)]{\mid} \card{\tips[\treeRV(\ttot)]}{=} n)
\prob(\card{\tips[\treeRV(\ttot)]}{=} n)
\right).\label{eq:empirVar}
\end{align}

\section{Birth-death random trees}\label{sec:birth_death}
Birth-death processes are classical stochastic models for population
dynamics and species diversification, and have long been used to model
phylogenetic trees \citep{Kendall1948b,Tavare2025}. In this framework, lineages
evolve independently, each giving rise to new lineages or becoming extinct
with per-lineage speciation and extinction rates, usually denoted by
$\birth$ and $\death$, respectively. When these rates are constant through
time, the process is said to be time-homogeneous. In this case, its qualitative
long-term behavior depends on the comparison between the birth and death
rates, leading to the classical supercritical, critical and subcritical regimes,
corresponding respectively to $\birth>\death$, $\birth=\death$, and
$\birth<\death$ \citep[see e.g.][]{Sagitov2013}.

\subsection{Distribution of the MRCA age under the generalized birth-death model}
The distribution of the MRCA age has been studied for Yule trees, i.e. trees
generated by a constant-rate pure-birth process
\citep{Sagitov2012,Mulder2015,Bartoszek2014,Bartoszek2015}.

Here, diversification follows a generalized birth-death process in the sense
of \citet{Kendall1948a}, with time-dependent per-lineage speciation and
extinction rates $\birth(\ttot)$ and $\death(\ttot)$ at time
$\ttot$.

Our strategy for computing the distribution of the MRCA age at a given time
$\ttot$ in this setting is based on the following observations. First, by construction, the MRCA
ages at time $\ttot$ depend only on the lineages extant at $\ttot$ and on the
corresponding reconstructed tree (Section~\ref{sec:fixed:phylo}). Second,
conditionally on survival up to time $\ttot$, this reconstructed tree is the realization of a time-inhomogeneous pure-birth process derived from the initial generalized
birth-death process~\citep{Nee1994}. Third, a deterministic time change transforms
this process into a unit-rate Yule process~\citep{Kendall1948b, Tavare2025}.
Finally, although we could use previous methods to obtain the distribution of the MRCA
age at this stage, we propose an alternative approach that is more
self-contained and better suited to our derivations.

\subsubsection{Reconstructed process and time change}\label{sec:bd:dist:rec}
Following \citet{Kendall1948a}, the probability $\pSurvival(\tfir,\ttot)$ that a lineage alive at time $\tfir$ leaves some extant descendant at time $\ttot\geq \tfir$ under the generalized birth-death process with time-dependent rates $\birth$ and $\death$ is 
\begin{equation*}
\pSurvival(\tfir,\ttot)
= \frac{1}{1+\int_{\tfir}^{\ttot}\mexp{-\int_{\tfir}^{u}(\birth(v)-\death(v))\mder v}\death(u)\mder u}.
\end{equation*}

For any time $\ttot$, the reconstructed process at time $\ttot$ associated with a diversification process is the stochastic process whose realizations contain only the lineages that have extant descendants at time $\ttot$. In particular, if diversification follows a generalized birth-death process, then, conditionally on non-extinction at time $\ttot$, the reconstructed process is a time-inhomogeneous pure-birth process with rate
\[
\rebirth[\ttot](\tfir)=\birth(\tfir)\pSurvival(\tfir,\ttot),
\]
for all $0\leq \tfir\leq \ttot$ \citep{Nee1994}.

Note that, by construction, if $\tree$ is the tree associated with a realization of a diversification process, then the tree $\tree(\ttot)$ defined in Section~\ref{sec:fixed:phylo} is associated with the corresponding realization of the reconstructed process at time $\ttot$.

Let us define the time change $\tfir \rightarrow \ftime(\tfir)$ with
\[
\ftime(\tfir)=\int_{0}^{\tfir}\rebirth[\ttot](w)\mder w.
\]
Under this deterministic time change, the reconstructed process becomes a Yule process with constant birth rate $1$
\citep{Kendall1948b, Tavare2025}.

\subsubsection{MRCA age distribution}
We are now ready to compute the distribution of the MRCA age of two tips
uniformly chosen among those extant at a given time. After the deterministic
time change, these tips can be regarded as tips of a unit-rate Yule tree
observed at a fixed height.

We first briefly recall two previous approaches. In
\citet{Sagitov2012}, the point-process representation of the Yule process,
together with results from \citet{Stadler2009}, is used to show that,
conditionally on observing $n$ tips, the MRCA age distribution of two uniformly chosen tips can be written as a
mixture, for $k=1,\ldots,n-1$, of the distributions of the maximum of $k$
independent and identically distributed speciation times, weighted by the
probabilities that two tips uniformly sampled among $n$ linearly ordered tips
are at distance $k$ in this order.

The approach of \citet{Mulder2015} also relies on results from
\citet{Stadler2009}. In this work, the MRCA age distribution is written as a
mixture, for $k=1,\ldots,n-1$, of the distributions of the $k$-th speciation
time, weighted by the probability that two randomly chosen tips coalesce at
that speciation time.

Although both approaches of \citet{Sagitov2012,Mulder2015} could be adapted
for the present purpose, we shall use an alternative approach for the following
reasons. First, applying these previous results in our setting would require
nontrivial adaptations. In \citet{Sagitov2012}, the expressions provided involve
quantities conditioned on the observed number of tips alone rather than on the
final time. The results of \citet{Mulder2015} apply to crown trees, whereas the
birth-death processes considered here start from a single lineage and therefore
generate stem trees. Second, the approach presented below relies only on transition
probabilities in the Yule process and is therefore more self-contained.
Finally, it yields convenient expressions for our derivations.

For all $0\le \tfir \le \ttot$, let
$\effRec[\treeRV(\ttot)](\tfir)$ denote the number of lineages at time $\tfir$
in the reconstructed process associated with the generalized birth-death
process. Equivalently, after the deterministic time change $\ftime$, it is the
number of lineages at time $\ftime(\tfir)$ in the corresponding unit-rate Yule
process. By construction, $\effRec[\treeRV(\ttot)](\ttot)$ is equal to the number of extant lineages at time $\ttot$ in the original diversification process, namely, $\effRec[\treeRV(\ttot)](\ttot)=\card{\tips[\treeRV(\ttot)]}$.

\begin{proposition}[adapted from Theorem~4.1 in \citet{Stadler2009}]\label{prop:coal}
Let $\treeRV$ be a pure-birth, lineage-homogeneous random tree. Let
$\tfir\leq \ttot$ be two times within the temporal span of $\treeRV$, and let
$m$ and $n$ be positive integers such that $m\leq n$ and $n\geq 2$. Conditional on observing $m$ lineages at time $\tfir$ and $n$ lineages at time $\ttot$, the probability that the MRCA age of two lineages sampled uniformly among the $n$ lineages present at time $\ttot$ is at most $\tfir$ is 
\[\prob\left(\mrcaAgeRVU[\treeRV(\ttot)]\leq \tfir \mid \effRec[\treeRV(\ttot)](\tfir) = m , \effRec[\treeRV(\ttot)](\ttot) = n\right)
=\left(1-\frac{2}{m+1}\right)\left(1+\frac{2}{n-1}\right).
\]
\end{proposition}

Under a generalized birth-death process, we write $\prob_{\ttot}$ for the law
conditional on survival of the original process up to time $\ttot$:
\[
\prob_{\ttot}(\cdot)
=
\prob(\cdot \mid \effRec[\treeRV(\ttot)](\ttot)\geq 1).
\]
Conditionally on survival up to $\ttot$, the reconstructed process at $\ttot$ becomes, after the deterministic time change $\ftime$, 
a unit-rate Yule process. For $0\leq\tau'\leq\tau\leq\ttot$ and $n\ge m\ge 1$, its transition probability is \citep{Kendall1948b, Tavare2025}
\[
\prob_{\ttot}\left(\effRec[\treeRV(\ttot)](\tau)=n \mid \effRec[\treeRV(\ttot)](\tau')=m\right)
=
\binom{n-1}{m-1}\,\mexp{-m(\ftime(\tau)-\ftime(\tau'))}\,\left(1-\mexp{-(\ftime(\tau)-\ftime(\tau'))}\right)^{n-m}.
\]
In particular, starting from a single lineage at time $0$, we have for all $0\leq\tfir\leq\ttot$,
\[
\prob_{\ttot}\left(\effRec[\treeRV(\ttot)](\tfir)=m\right)=\mexp{-\ftime(\tfir)}\left(1-\mexp{-\ftime(\tfir)}\right)^{m-1}.
\]
Therefore, for $n\ge m\ge 1$ and $0\leq\tfir\leq\ttot$,
\begin{align}
\prob_{\ttot}\left(\effRec[\treeRV(\ttot)](\tfir)=m,\ \effRec[\treeRV(\ttot)](\ttot)=n\right)
&=\prob_{\ttot}\left(\effRec[\treeRV(\ttot)](\tfir)=m\right)\,
 \prob_{\ttot}\left(\effRec[\treeRV(\ttot)](\ttot)=n\mid \effRec[\treeRV(\ttot)](\tfir)=m\right)\notag\\
&=
\mexp{-\ftime(\tfir)}\left(1-\mexp{-\ftime(\tfir)}\right)^{m-1}
\binom{n-1}{m-1}
\mexp{-m(\ftime(\ttot)-\ftime(\tfir))}\left(1-\mexp{-(\ftime(\ttot)-\ftime(\tfir))}\right)^{n-m}.\label{eq:trans}
\end{align}

For all integers $1\leq m \leq n$, combining Proposition~\ref{prop:coal} with the transition probabilities of the reconstructed process yields the joint probability

\begin{multline*}
\prob_{\ttot}\left(\mrcaAgeRVU[\treeRV(\ttot)]\leq \tfir, \effRec[\treeRV(\ttot)](\tfir) = m , \effRec[\treeRV(\ttot)](\ttot) = n\right)\\
=\prob_{\ttot}\left(\mrcaAgeRVU[\treeRV(\ttot)]\leq \tfir \mid \effRec[\treeRV(\ttot)](\tfir) = m , \effRec[\treeRV(\ttot)](\ttot) = n\right)\prob_{\ttot}\left(\effRec[\treeRV(\ttot)](\tfir)=m,\ \effRec[\treeRV(\ttot)](\ttot)=n\right)
\end{multline*}

This gives the contribution of each pair $(m,n)$ to the event $\mrcaAgeRVU[\treeRV(\ttot)]\leq\tfir$. For a given integer $n\geq 2$, summing these contributions over all integers $1\leq m \leq n$, and normalizing, yields the following explicit expression for the
distribution of the MRCA age conditional on observing $n$ tips.
\begin{proposition}\label{prop:mrcaDistCond}
Let $\treeRV$ be the random tree generated by a generalized birth-death processs,
and let $\ftime$ be the associated time-change function. Then, for all $n\geq 2$ and every $0<\tfir\le \ttot$, the distribution function of
$\mrcaAgeRVU[\treeRV(\ttot)]$ conditional on observing exactly $n$ lineages at
time $\ttot$, namely, $\distri{\mrcaAgeRVU[\treeRV(\ttot)]\mid_{\overset{=}{n}}}(\tfir) = \prob_{\ttot}\left( \mrcaAgeRVU[\treeRV(\ttot)]\le \tfir \mid \effRec[\treeRV(\ttot)](\ttot)=n \right)$, is
\[
\distri{\mrcaAgeRVU[\treeRV(\ttot)]\mid_{\overset{=}{n}}}(\tfir) = 1 - \frac{2}{n}\left( \frac{ \mexp{\ftime(\ttot)}-\mexp{\ftime(\tfir)} }{ \mexp{\ftime(\tfir)}-1 } \right)\left(1 - \frac{1}{n-1}\left( \frac{ \mexp{\ftime(\ttot)}-\mexp{\ftime(\tfir)} }{ \mexp{\ftime(\tfir)}-1 } \right) \left[ 1- \left( \frac{ \mexp{\ftime(\ttot)}-\mexp{\ftime(\tfir)} }{ \mexp{\ftime(\ttot)}-1 } \right)^{n-1} \right]\right). 
\]
Moreover, by continuity, $\distri{\mrcaAgeRVU[\treeRV(\ttot)]\mid_{\overset{=}{n}}}(0) = 0$.

If $\ftime$ is differentiable at $\tfir$, the associated conditional density is 
\[\densi{\mrcaAgeRVU[\treeRV(\ttot)]\mid_{\overset{=}{n}}}(\tfir) = \ftime'(\tfir) \frac{ 2\mexp{\ftime(\tfir)} \left(\mexp{\ftime(\ttot)}-1\right) }{ n\left(\mexp{\ftime(\tfir)}-1\right)^2 } \left[ 1 - \frac{ 2\left(\mexp{\ftime(\ttot)}-\mexp{\ftime(\tfir)}\right) }{ (n-1)\left(\mexp{\ftime(\tfir)}-1\right) } \left( 1- \left( \frac{ \mexp{\ftime(\ttot)}-\mexp{\ftime(\tfir)} }{ \mexp{\ftime(\ttot)}-1 } \right)^{n-1} \right) + \left( \frac{ \mexp{\ftime(\ttot)}-\mexp{\ftime(\tfir)} }{ \mexp{\ftime(\ttot)}-1 } \right)^n \right].
\]
\end{proposition}

The conditional expectation of $\mrcaAgeRVU[\treeRV(\ttot)]$ can be obtained by integrating its survival function, that is,

\begin{multline*}
\esp\left( \mrcaAgeRVU[\treeRV(\ttot)] \mid \effRec[\treeRV(\ttot)](\ttot)=n \right) = \int_{0}^{\ttot} (1-\distri{\mrcaAgeRVU[\treeRV(\ttot)]\mid_{\overset{=}{n}}}(\tfir))\mder \tfir\\
=
\frac{2}{n}\int_{0}^{\ttot}\left( \frac{ \mexp{\ftime(\ttot)}-\mexp{\ftime(\tfir)} }{ \mexp{\ftime(\tfir)}-1 } \right)\left(1 - \frac{1}{n-1}\left( \frac{ \mexp{\ftime(\ttot)}-\mexp{\ftime(\tfir)} }{ \mexp{\ftime(\tfir)}-1 } \right) \left[ 1- \left( \frac{ \mexp{\ftime(\ttot)}-\mexp{\ftime(\tfir)} }{ \mexp{\ftime(\ttot)}-1 } \right)^{n-1} \right]\right)\mder \tfir
\end{multline*}

Summing $\prob_{\ttot}\left(\mrcaAgeRVU[\treeRV(\ttot)]\leq \tfir, \effRec[\treeRV(\ttot)](\tfir) = m , \effRec[\treeRV(\ttot)](\ttot) = n\right)$ over all integers $n\geq 2$ and all $1\leq m \leq n$, and normalizing, yields the following proposition.

\begin{proposition}\label{prop:mrcaDist}
Let $\treeRV$ be the random tree generated by a generalized birth-death processs,
and let $\ftime$ be the associated time-change function. Then, for every $0<\tfir\le \ttot$, the distribution function of
$\mrcaAgeRVU[\treeRV(\ttot)]$ conditional on observing at least two lineages at
time $\ttot$, namely, $\distri{\mrcaAgeRVU[\treeRV(\ttot)]\mid_{\overset{\geq}{2}}}(\tfir) = \prob_{\ttot}\left( \mrcaAgeRVU[\treeRV(\ttot)]\le \tfir \mid \effRec[\treeRV(\ttot)](\ttot)\geq 2 \right)$, is
\[
\distri{\mrcaAgeRVU[\treeRV(\ttot)]\mid_{\overset{\geq}{2}}}(\tfir)
=
1-2\frac{(\mexp{\ftime(\ttot)}-\mexp{\ftime(\tfir)})\left(1-\mexp{\ftime(\tfir)}+\ftime(\tfir)\mexp{\ftime(\tfir)}\right)}{(\mexp{\ftime(\ttot)}-1)\left(\mexp{\ftime(\tfir)}-1\right)^{2}}.
\]
Moreover, by continuity, $\distri{\mrcaAgeRVU[\treeRV(\ttot)]\mid_{\overset{\geq}{2}}}(0) = 0$.

If $\ftime$ is differentiable at $\tfir$, the associated conditional density is
\[
\densi{\mrcaAgeRVU[\treeRV(\ttot)]\mid_{\overset{\geq}{2}}}(\tfir) = \frac{
2\ftime'(\tfir)\mexp{\ftime(\tfir)}
}{
\left(\mexp{\ftime(\ttot)}-1\right)
\left(\mexp{\ftime(\tfir)}-1\right)^3
}
\left[
\left(\mexp{\ftime(\ttot)}-\mexp{\ftime(\tfir)}\right)
\left(1+\ftime(\tfir)-\mexp{\ftime(\tfir)}\right)
+
\left(\mexp{\ftime(\ttot)}-1\right)
\left(1+\ftime(\tfir)\mexp{\ftime(\tfir)}-\mexp{\ftime(\tfir)}\right)
\right].
\]
\end{proposition}

The conditional expectation of $\mrcaAgeRVU[\treeRV(\ttot)]$ can be obtained again by integrating its survival function, that is,
\begin{equation*}
\esp\left( \mrcaAgeRVU[\treeRV(\ttot)] \mid \effRec[\treeRV(\ttot)](\ttot)\geq 2 \right) = \int_{0}^{\ttot} (1-\distri{\mrcaAgeRVU[\treeRV(\ttot)]\mid_{\overset{\geq}{2}}}(\tfir))\mder \tfir
=\frac{2}{\mexp{\ftime(\ttot)}-1}\int_{0}^{\ttot}\frac{(\mexp{\ftime(\ttot)}-\mexp{\ftime(\tfir)})\left(1-\mexp{\ftime(\tfir)}+\ftime(\tfir)\mexp{\ftime(\tfir)}\right)}{\left(\mexp{\ftime(\tfir)}-1\right)^{2}}\mder \tfir
\end{equation*}

\begin{figure}
\centerline{\includegraphics[width=0.66\textwidth]{"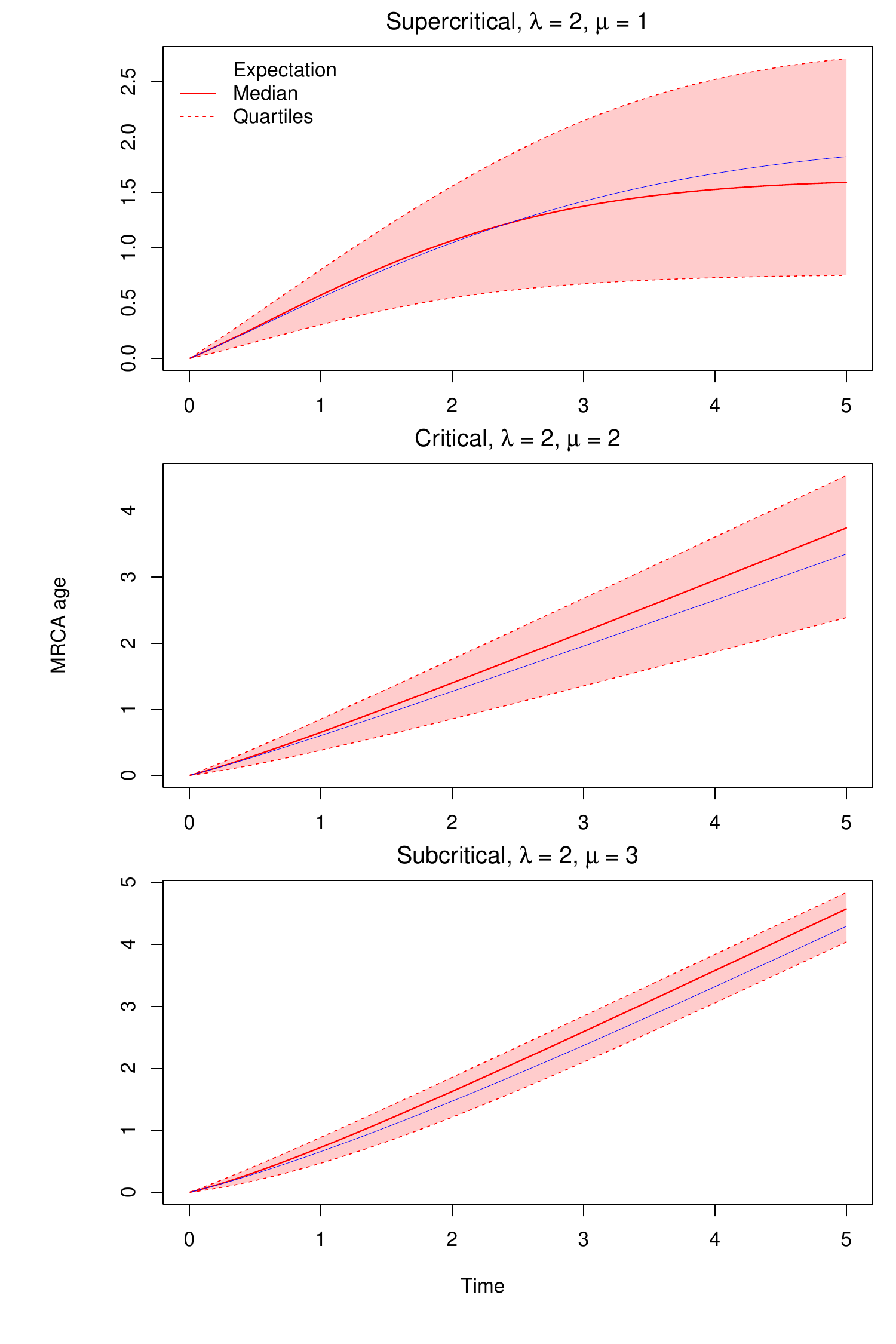"}}
\caption{Dynamics of the MRCA age distribution for two uniformly sampled extant tips of a birth-death tree, conditional on observing at least two extant lineages, computed from Proposition~ \ref{prop:mrcaDist}.}\label{fig:mrcaConstRates}
\end{figure}

Figure~\ref{fig:mrcaConstRates} presents the dynamics of the distribution of the MRCA age of two uniformly sampled extant tips for the three constant birth-death regimes, conditional on observing at least two lineages. Overall, the spread of the distribution increases with time, whereas the behavior of the expectation depends strongly on the regime considered: it appears to increase linearly in both the critical and subcritical regimes, whereas its slope decreases over time in the supercritical case.

\subsection{General expressions for the variance of the empirical mean and for the expected empirical variance}

Since the generalized birth-death processes are lineage-homogeneous, the
results of Section~\ref{sec:linHom} apply. In particular, for $n\geq 2$, the relation between
the expected MRCA age and the variance of the empirical mean of a Brownian trait
gives
\[
\variance(\meanTrait[\treeRV(\ttot)]\mid \card{\tips[\treeRV(\ttot)]}=n)
=
\frac{\sigma^2}{n}\left(\ttot+(n-1)\esp(\mrcaAgeRVU[\treeRV(\ttot)]\mid \card{\tips[\treeRV(\ttot)]}=n)\right).
\]

By construction, we have $\card{\tips[\treeRV(\ttot)]}=\effRec[\treeRV(\ttot)](\ttot)$ and, for any event $B$,
$\prob(B\mid \card{\tips[\treeRV(\ttot)]}\ge 1)=\prob_{\ttot}(B)$.
It follows that 
\begin{align*}
\variance(\meanTrait[\treeRV(\ttot)]\mid \card{\tips[\treeRV(\ttot)]}\geq 1)
 &= {\sum_{n=1}^{\infty}\variance(\meanTrait[\treeRV(\ttot)]\mid \card{\tips[\treeRV(\ttot)]}=n)\prob_{\ttot}(\effRec[\treeRV(\ttot)](\ttot)=n)}\\
 &={\sigma^{2}\ttot\prob_{\ttot}(\effRec[\treeRV(\ttot)](\ttot)=1)+\sum_{n=2}^{\infty}\frac{\sigma^2}{n}\left(\ttot+(n-1)\esp(\mrcaAgeRVU[\treeRV(\ttot)]\mid \card{\tips[\treeRV(\ttot)]}=n)\right)\prob_{\ttot}(\effRec[\treeRV(\ttot)](\ttot)=n)}
 \end{align*}
Computing the sum above leads to the following result.
\begin{proposition}\label{prop:variance}
Let $\treeRV$ be the random tree generated by a generalized birth-death processs,
and let $\ftime$ be the associated time-change function. The variance of the empirical mean at time $\ttot$ of a trait evolving along $\treeRV$ according to a Brownian motion model with variance parameter $\sigma^{2}$, conditional on survival at $\ttot$, is given by
\begin{multline*}
\variance(\meanTrait[\treeRV(\ttot)]\mid \effRec[\treeRV(\ttot)](\ttot)\geq 1)\\
=
\sigma^2\left(-\frac{\ttot\ftime(\ttot)}{\mexp{\ftime(\ttot)}-1}+2\int_0^{\ttot}\left(
\frac{\ftime(\ttot)}{\mexp{\ftime(\tfir)}-1}+\frac{\mexp{\ftime(\ttot)}-\mexp{\ftime(\tfir)}}{(\mexp{\ftime(\tfir)}-1)^2}
\left[
\operatorname{Li}_2\left(1-\mexp{-(\ftime(\ttot)-\ftime(\tfir))}\right)
-
\operatorname{Li}_2\left(1-\mexp{-\ftime(\ttot)}\right)
\right]\right)\mder\tfir\right),
\end{multline*}
where $\operatorname{Li}_2$ denotes the dilogarithm function (Appendix~\ref{app:dilog}).
\end{proposition}

\begin{figure}
\centerline{\includegraphics[width=0.66\textwidth]{"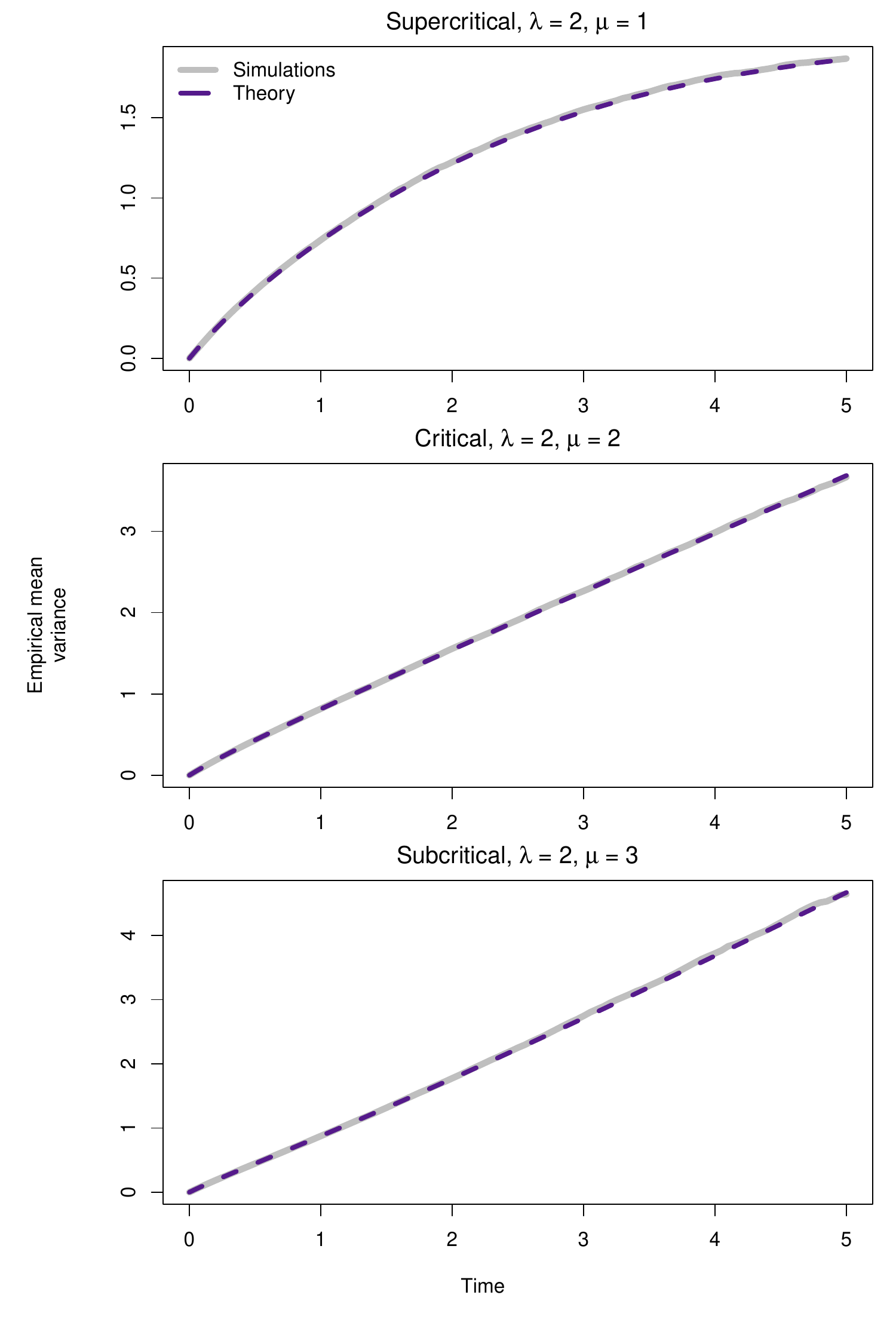"}}
\caption{Variance of the empirical mean of a standard Brownian trait evolving on birth-death trees conditional on survival, computed from Proposition~\ref{prop:variance} and estimated from $5\times 10^4$, $5\times 10^5$ and $5\times 10^6$ simulations for
the supercritical, critical and subcritical regimes, respectively.}
\label{fig:meanVarianceConstRates}
\end{figure}

Figure~\ref{fig:meanVarianceConstRates} illustrates the behavior of the variance
of the empirical mean under the three constant birth-death regimes. Its dynamics
is similar to that of the expected MRCA age displayed in
Figure~\ref{fig:mrcaConstRates}: it grows approximately linearly in the critical
and subcritical regimes, whereas its slope decreases over time in the
supercritical regime.

The theoretical values, computed from Proposition~\ref{prop:variance}, are shown
together with estimates obtained from $5\times 10^4$, $5\times 10^5$ and $5\times 10^6$ simulations for
the supercritical, critical and subcritical regimes, respectively. The number of
simulations was adjusted across regimes to compensate for the different
extinction probabilities and to obtain sufficiently stable Monte Carlo estimates
in all three cases. For each replicate, we first simulate a birth-death tree
under the considered rates, and then simulate a standard Brownian trait along
this tree. At each point of a fixed time grid, and for each replicate that has
not gone extinct, we compute the empirical mean of the trait values across the
lineages extant at that time. The plotted simulation estimate is the sample
variance, across all non-extinct replicates at that time, of the resulting
empirical means.

We now return to Equation~\ref{eq:empirVar} in order to evaluate the expected empirical variance. 
\[
\sum_{n=2}^{\infty}\esp(\mrcaAgeRVU[\treeRV(\ttot)]\mid \card{\tips[\treeRV(\ttot)]}=n)\prob(\card{\tips[\treeRV(\ttot)]}=n)
=
\prob(\card{\tips[\treeRV(\ttot)]}\ge 2)\,
\esp(\mrcaAgeRVU[\treeRV(\ttot)]\mid \card{\tips[\treeRV(\ttot)]}\ge 2).
\]

Since the event $\card{\tips[\treeRV(\ttot)]}\geq 2$ implies survival up to time $\ttot$ and $\card{\tips[\treeRV(\ttot)]}=\effRec[\treeRV(\ttot)](\ttot)$, we have for any event $B$,
\[
\prob(B\mid \card{\tips[\treeRV(\ttot)]}\ge 2)=\prob_{\ttot}(B\mid \effRec[\treeRV(\ttot)](\ttot)\ge 2).
\]

Moreover, since $0\le \mrcaAgeRVU[\treeRV(\ttot)]\le \ttot$, we have
\[
\esp(\mrcaAgeRVU[\treeRV(\ttot)]\mid \card{\tips[\treeRV(\ttot)]}\ge 2)
=
\int_0^{\ttot}\left(1-\prob_{\ttot}\left(\mrcaAgeRVU[\treeRV(\ttot)]\leq \tfir \mid \effRec[\treeRV(\ttot)](\ttot)\ge 2\right)\right)\mder\tfir.
\]
Therefore, using Equation~\ref{eq:empirVar}, we obtain
\[
\esp\left(\mavar{\treeRV(\ttot)}\right)
=
\sigma^{2}\prob(\card{\tips[\treeRV(\ttot)]}\ge 2)\int_0^{\ttot}\prob_{\ttot}\left(\mrcaAgeRVU[\treeRV(\ttot)]\leq \tfir \mid \effRec[\treeRV(\ttot)](\ttot)\ge 2\right)\mder\tfir.
\]

We have that 
\begin{align*}
\prob(\card{\tips[\treeRV(\ttot)]}\ge 2)
&=\prob(\card{\tips[\treeRV(\ttot)]}>0)\prob_{\ttot}(\effRec[\treeRV(\ttot)](\ttot)\ge 2)
\\&=
\pSurvival(0,\ttot)\left(1-\mexp{-\ftime(\ttot)}\right)
\end{align*}
Putting these observations together yields directly the following proposition.
\begin{proposition}\label{prop:expect}
Let $\treeRV$ be the random tree generated by a generalized birth-death processs,
and let $\ftime$ be the associated time-change function. The expected empirical variance at time $\ttot$ of a trait evolving along $\treeRV$ according to a Brownian motion model with variance parameter $\sigma^{2}$ is
\[
\esp\left(\mavar{\treeRV(\ttot)}\right)
=
\sigma^{2}\pSurvival(0,\ttot)(1-\mexp{-\ftime(\ttot)})\left(\ttot -\frac{2}{\mexp{\ftime(\ttot)}-1}\int_0^{\ttot}\frac{(\mexp{\ftime(\ttot)}-\mexp{\ftime(\tfir)})\left(1-\mexp{\ftime(\tfir)}+\ftime(\tfir)\mexp{\ftime(\tfir)}\right)}{\left(\mexp{\ftime(\tfir)}-1\right)^{2}}\mder\tfir\right).
\]
\end{proposition}

Note that the expected empirical variance in Proposition~\ref{prop:expect} is not conditioned on survival up to time $\ttot$; it also accounts for trees that became extinct before $\ttot$. Conditioning the expected empirical variance on survival up to $\ttot$ is
performed simply by removing the factor $\pSurvival(0,\ttot)$ from the above
expression.
\begin{figure}
\centerline{\includegraphics[width=0.66\textwidth]{"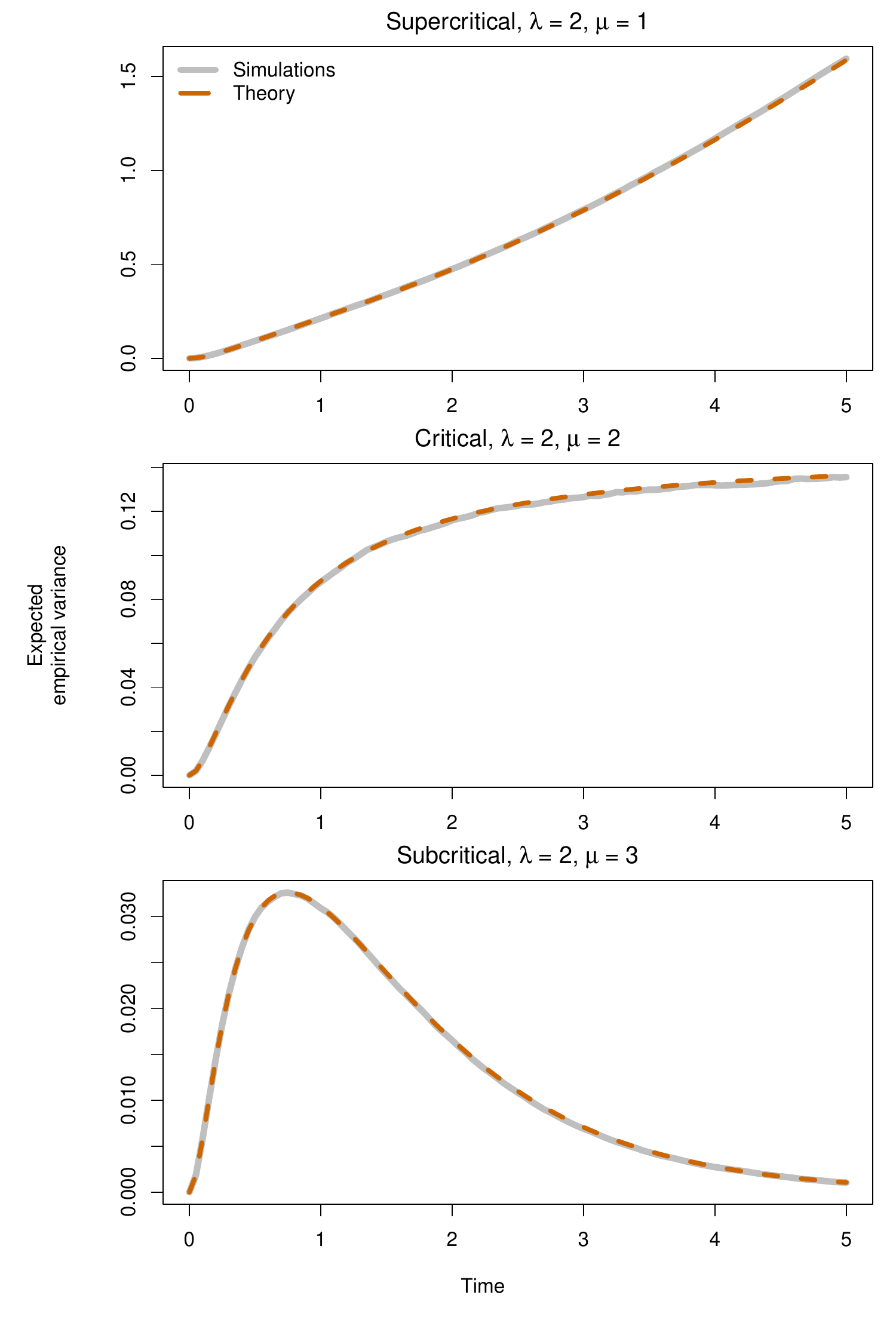"}}
\caption{Unconditional expected empirical variance of a standard Brownian trait evolving on birth-death trees, computed from Proposition~\ref{prop:expect} and estimated from $5\times 10^4$, $5\times 10^5$ and $5\times 10^6$ simulations for
the supercritical, critical and subcritical regimes, respectively.}
\label{fig:empiricalVarianceUncond}
\end{figure}

\begin{figure}
\centerline{\includegraphics[width=0.66\textwidth]{"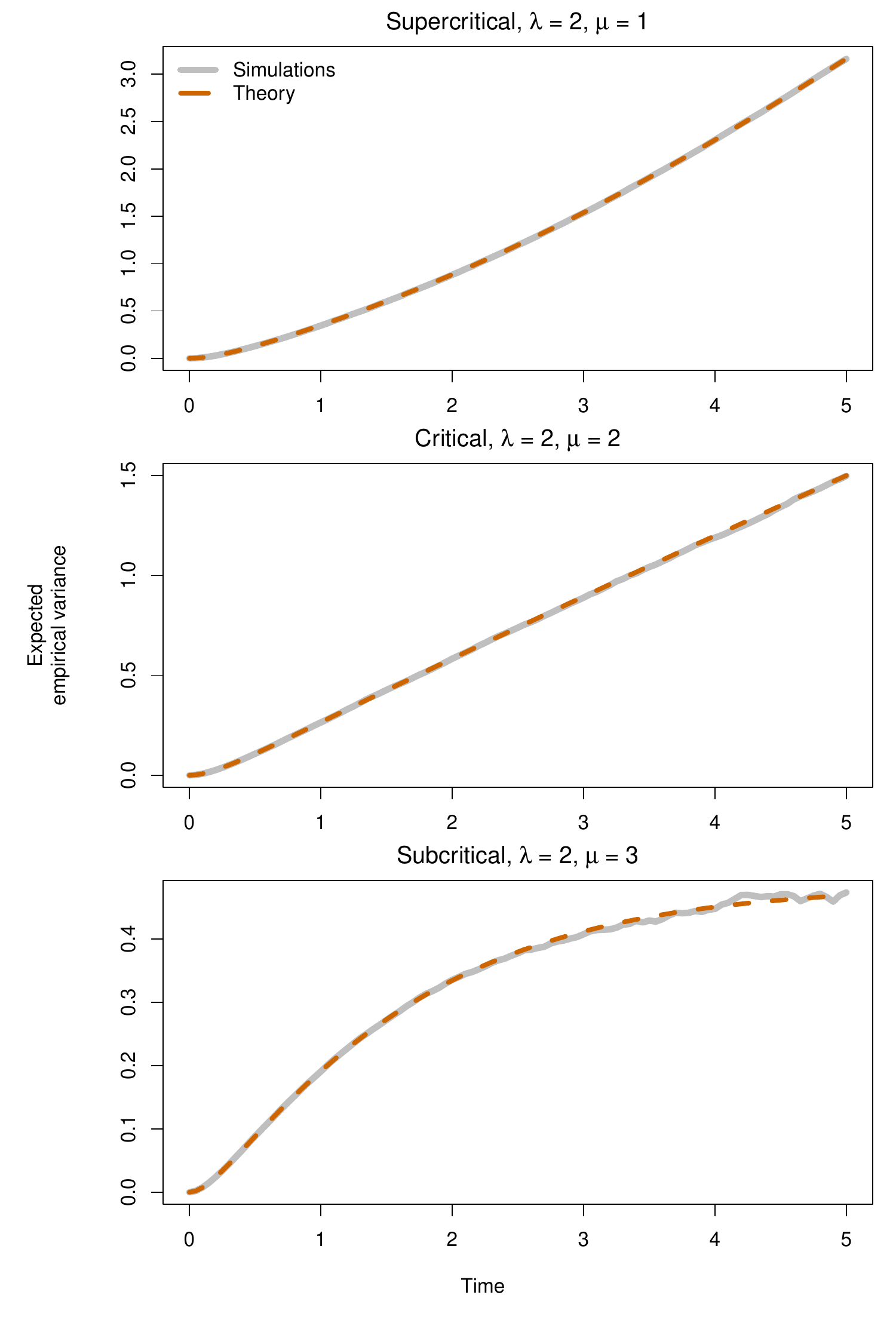"}}
\caption{Expected empirical variance of a standard Brownian trait evolving on birth-death trees, conditional on survival, computed from Proposition~\ref{prop:expect} and estimated from $5\times 10^4$, $5\times 10^5$ and $5\times 10^6$ simulations for
the supercritical, critical and subcritical regimes, respectively.}
\label{fig:empiricalVarianceCond}
\end{figure}
Figures~\ref{fig:empiricalVarianceUncond} and~\ref{fig:empiricalVarianceCond}
present the dynamics of the expected empirical variance under the three regimes
of the constant birth-death process, respectively without conditioning and
conditionally on survival.

The simulation protocol is analogous to that used for
Figure~\ref{fig:meanVarianceConstRates}. For each replicate, we first simulate a
birth-death tree under the considered rates, then simulate a standard Brownian
trait along this tree. At each point of a fixed time grid, we compute the
empirical variance of the trait values across the extant lineages, with the
convention that this variance is zero whenever fewer than two lineages are
present. The plotted simulation estimate is then the average of these empirical
variances across replicates. In the unconditional case, all replicates contribute
to this average, including those that are extinct at the considered time. In the
survival-conditioned case, only non-extinct replicates contribute, but replicates
with a single extant lineage still contribute zero.

In the unconditional case, the three regimes lead to markedly different
behaviors. The expected empirical variance grows approximately linearly in the
supercritical regime. In the critical regime, it increases with a decreasing
slope, consistently with convergence to a finite limit. In the subcritical
regime, after a short initial increase, it decreases toward zero as extinction
becomes dominant (Figure~\ref{fig:empiricalVarianceUncond}).

Conditioning on survival changes the picture. The expected empirical variance
still grows approximately linearly in the supercritical regime, and it also grows
over the displayed time range in the critical regime. In the subcritical regime,
it no longer decreases toward zero, but instead appears to approach a positive
limiting value. Since the number of non-extinct replicates decreases through
time, especially in the subcritical regime, the simulation estimate is less
stable at large times in this regime (Figure~\ref{fig:empiricalVarianceCond}).

\begin{figure}
\centerline{\includegraphics[width=0.66\textwidth]{"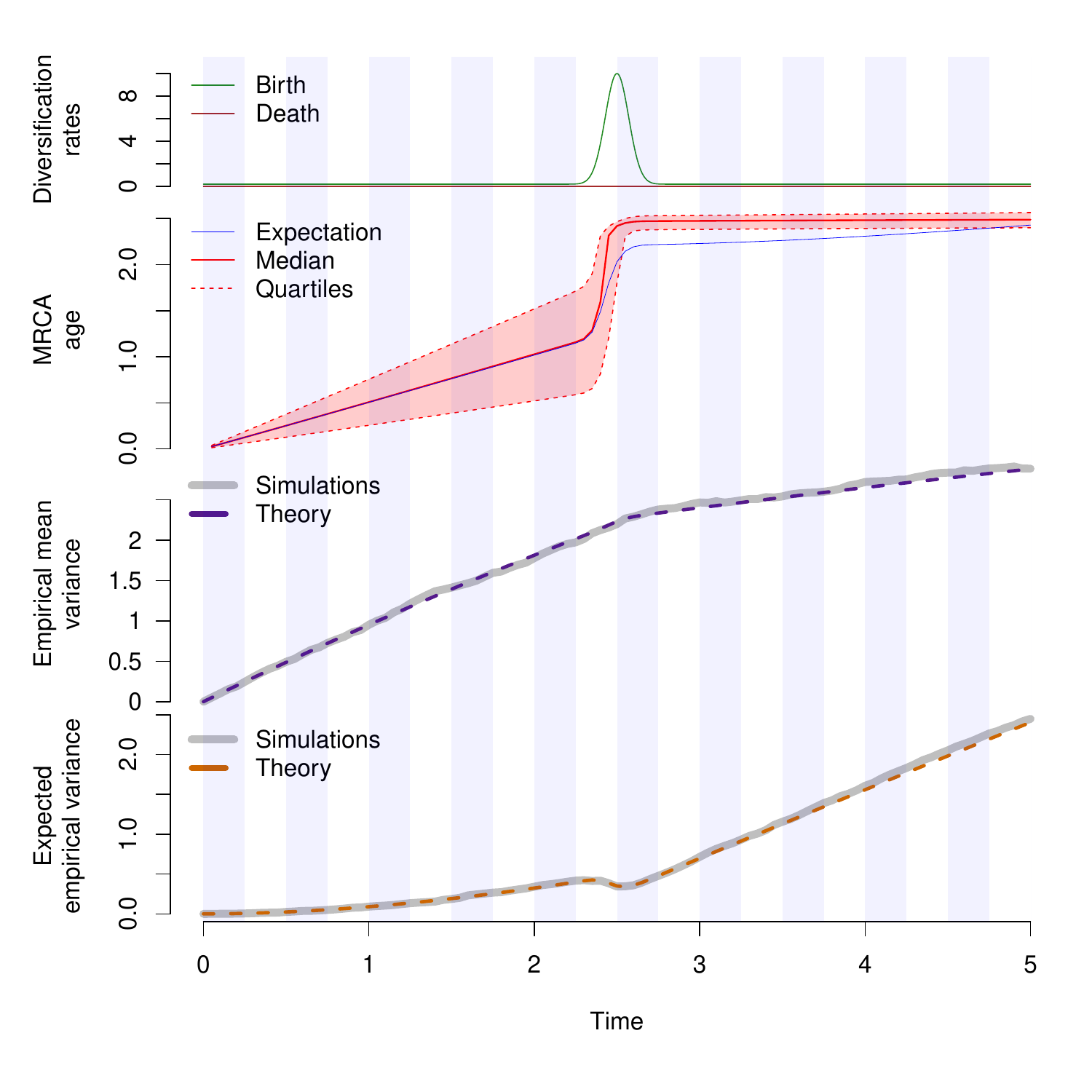"}}
\caption{
Dynamics under a time-dependent pure-birth process.
Top panel: birth rate, with a peak at time 2.5, and constant zero death rate.
Second panel: MRCA age distribution for two uniformly sampled extant tips.
Third panel: variance of the empirical mean of a standard Brownian trait evolving on the corresponding birth-death trees, computed from Proposition~\ref{prop:variance} and estimated from 1000 simulations.
Bottom panel: expected empirical variance of the same trait, computed from Proposition~\ref{prop:expect} and estimated from 1000 simulations.
}\label{fig:variableRatesPaths}
\end{figure}

Figure~\ref{fig:variableRatesPaths} illustrates the general formulas obtained so far on a time-dependent pure-birth example. It displays, in a single figure, the diversification rates, the MRCA age distribution, the variance of the empirical mean, and the expected empirical variance. We consider a time-inhomogeneous pure-birth process with a burst of speciation around time 2.5.

The effect of the burst on the MRCA age distribution is striking: all quartiles, including the median, increase abruptly during the burst and then remain almost constant, with a reduced interquartile range. This indicates that, after the burst, a large fraction of pairs of extant lineages have an MRCA associated with a speciation event occurring during this short period. The near constancy of the quartiles after the burst, while the expectation continues to increase at a lower rate, may reflect the fact that the low post-burst speciation rate produces too few new lineages to substantially modify the bulk of the distribution over short time scales. Both the variance of the empirical mean and the expected empirical variance show visible changes in slope around the burst, with an additional inflection in the latter case.

\subsection{Expected empirical variance in the constant birth-death case}
We restrict the constant-rate analysis to the expected empirical variance, both
because it leads to simpler formulas and because it appears to be of broader
biological interest.

Proposition~\ref{prop:expect} provides, at any time, a general integral expression for the expected empirical variance of trait values evolving under Brownian motion on a random tree generated by a generalized birth-death process, and therefore allows for numerical evaluation. 
In the constant-rate case, this expression yields closed-form formulas involving the dilogarithm function~\citep{Morris1979}.
In this section, $\birth$ and $\death$ are no longer functions of time but
constants. Following~\citet{Kendall1948a}, the
expressions introduced in Section~\ref{sec:bd:dist:rec} become,
in the non-critical case, i.e., $\birth\neq\death$,
\[
\pSurvival(\tfir,\ttot) = \frac{\birth-\death}{\birth-\death\mexp{-(\birth-\death)(\ttot-\tfir)}} \;,\;\rebirth[\ttot](\tfir) =\frac{\birth(\birth-\death)}{\birth-\death\mexp{-(\birth-\death)(\ttot-\tfir)}} \;,\;\ftime(\tfir) = (\birth-\death)\tfir-\log\left(\frac{\birth-\death \mexp{-(\birth-\death)(\ttot-\tfir)}}{\birth-\death \mexp{-(\birth-\death)\ttot}}\right) ,
\]
whereas, in the critical case, i.e., $\birth=\death$, they become
\[
\pSurvival(\tfir,\ttot) = \frac{1}{1+\birth(\ttot-\tfir)} \;,\;\rebirth[\ttot](\tfir) =\frac{\birth}{1+\birth(\ttot-\tfir)} \;,\;\ftime(\tfir) = \log\left(\frac{1+\birth\ttot}{1+\birth(\ttot-\tfir)}\right).
\]

Note that substituting the constant-rate expression for $\ftime$ in
Propositions~\ref{prop:mrcaDistCond} and~\ref{prop:mrcaDist} directly yields
closed-form expressions for the corresponding distribution and density functions of the
MRCA age. For instance, in the non-critical case $\birth\neq\death$, for
$0<\tfir\leq\ttot$, we obtain after simplification
\begin{multline*}
\distri{\mrcaAgeRVU[\treeRV(\ttot)]\mid_{\overset{\geq}{2}}}(\tfir)
=
1-
2\frac{
(\birth-\death \mexp{-(\birth-\death)\ttot})
(\mexp{-(\birth-\death)\tfir}-\mexp{-(\birth-\death)\ttot})
}{
\birth^{2}
(1-\mexp{-(\birth-\death)\ttot})
(1-\mexp{-(\birth-\death)\tfir})^{2}
}
\\
\times
\left[
-\birth(1-\mexp{-(\birth-\death)\tfir})
+
(\birth-\death \mexp{-(\birth-\death)\ttot})
\left(
(\birth-\death)\tfir
-
\log\left(
\frac{
\birth-\death \mexp{-(\birth-\death)(\ttot-\tfir)}
}{
\birth-\death \mexp{-(\birth-\death)\ttot}
}
\right)
\right)
\right].
\end{multline*}

\subsubsection{Closed expressions}

\begin{proposition}\label{prop:closednon}
Assume that diversification follows a birth-death process with constant speciation and extinction rates $\birth$ and $\death$, with $\birth \neq \death$, and let $\treeRV$ be the random tree associated with the process. The expected empirical variance at time $\ttot$ of a trait evolving along $\treeRV$ according to a Brownian motion model with variance parameter $\sigma^{2}$ is
\begin{multline*}
\esp(\mavar{\treeRV(\ttot)})
=
\frac{\sigma^2(\birth-\death)}{\birth-\death\mexp{-(\birth-\death)\ttot}}
\Bigg[
\ttot\left(
\frac{\birth\left(1-\mexp{-(\birth-\death)\ttot}\right)}{\birth-\death \mexp{-(\birth-\death)\ttot}}
-
\frac{2\death}{\birth}\mexp{-(\birth-\death)\ttot}
\right)
-
\frac{2\left(1-\mexp{-(\birth-\death)\ttot}\right)}{\birth-\death}
\\+
\frac{2\mexp{-(\birth-\death)\ttot}}{\birth}
\log\left(
\frac{\birth-\death \mexp{-(\birth-\death)\ttot}}{\birth-\death}
\right)
-
\frac{2\mexp{-(\birth-\death)\ttot}\left(\birth-\death \mexp{-(\birth-\death)\ttot}\right)}{\birth(\birth-\death)}
\left(
\operatorname{Li}_2\left(1-\mexp{(\birth-\death)\ttot}\right)
-
\operatorname{Li}_2\left(
\frac{\death\left(1-\mexp{-(\birth-\death)\ttot}\right)}
{\birth-\death \mexp{-(\birth-\death)\ttot}}
\right)
\right)
\Bigg].
\end{multline*}
In particular in the Yule case, i.e., $\death=0$, this reduces to
\[
\esp(\mavar{\treeRV(\ttot)})
=
\sigma^2
\left[
\ttot\left(1-\mexp{-\birth \ttot}\right)
-\frac{2}{\birth}\left(1-\mexp{-\birth \ttot}\left(1-\operatorname{Li}_2\left(1-\mexp{\birth \ttot}\right)\right)\right)
\right].
\]
\end{proposition}

\begin{proposition}\label{prop:closedcrit}
Assume that diversification follows a critical birth-death process with constant speciation and extinction rate $\birth=\death$, and let $\treeRV$ be the random tree associated with the process. The expected empirical variance at time $\ttot$ of a trait evolving along $\treeRV$ according to a Brownian motion model with variance parameter $\sigma^{2}$ is
\[
\esp(\mavar{\treeRV(\ttot)})
=
\frac{\sigma^2}{1+\birth\ttot}
\left[
\frac{\birth \ttot^2}{1+\birth \ttot}
-4\ttot
+\frac{2}{\birth}\log(1+\birth \ttot)
+\frac{2(1+\birth \ttot)}{\birth}
\operatorname{Li}_2\left(\frac{\birth \ttot}{1+\birth \ttot}\right)
\right].
\]
\end{proposition}

\subsubsection{Asymptotic behavior}
\begin{proposition}\label{prop:asymp}
Assume that diversification follows a birth-death process with constant speciation and extinction rates $\birth$ and $\death$ and let $\treeRV$ be the random tree associated with the process. The expected empirical variance at time $\ttot$ of a trait evolving along $\treeRV$ according to a Brownian motion model with variance parameter $\sigma^{2}$ satisfies, as $\ttot\rightarrow\infty$,
\[
\esp\left(\mavar{\treeRV(\ttot)}\right)
=
\left\{
\begin{array}{ll}
\frac{\sigma^2}{\birth}\left((\birth-\death)\ttot-2\right)+o(1)
& \mbox{if $\birth>\death$,}\\[1ex]
\frac{\sigma^2}{\birth}\left(\frac{\pi^2}{3}-3\right)+o(1)
& \mbox{if $\birth=\death$,}\\[1ex]
\sigma^2 \Upsilon_{\birth,\death}
\mexp{-(\death-\birth)\ttot}
+
o\left(\mexp{-(\death-\birth)\ttot}\right)
& \mbox{if $\birth<\death$,}
\end{array}
\right.
\]
where
\[
\Upsilon_{\birth,\death}
=
\frac{1}{\death^{2}}\left(\frac{(\death-\birth)^{2}}{\birth}\log\left(\frac{\death-\birth}{\death}\right)+\death-\frac{\birth}{2}\right).
\]
\end{proposition}

\begin{corollary}\label{cor:asympcond}
Assume that diversification follows a birth-death process with constant
speciation and extinction rates $\birth$ and $\death$, and let $\treeRV$ be the
random tree associated with the process. The expected empirical variance at time
$\ttot$ of a trait evolving along $\treeRV$ according to a Brownian motion model with variance parameter $\sigma^{2}$, conditional on non-extinction at time $\ttot$, satisfies, as
$\ttot\rightarrow\infty$,
\[
\esp\left(
\mavar{\treeRV(\ttot)}
\mid \effRec[\treeRV(\ttot)](\ttot)\geq 1
\right)
=
\left\{
\begin{array}{ll}
\sigma^2\left(\ttot-\dfrac{2}{\birth-\death}\right)+o(1)
& \mbox{if $\birth>\death$,}\\[2ex]
\sigma^2\left(\frac{\pi^2}{3}-3\right)\left(\ttot+\frac{1}{\birth}\right)
+
o(1)
& \mbox{if $\birth=\death$,}\\[2ex]
\sigma^{2}\frac{\death }{\death-\birth}\Upsilon_{\birth,\death}+o(1)
& \mbox{if $\birth<\death$.}
\end{array}
\right.
\]
\end{corollary}
\section{Discussion}
This study takes a global perspective on Brownian trait evolution by focusing
on the empirical mean and empirical variance of trait values across extant
lineages through time, rather than on individual lineages themselves.
A first point emerging from the results is that the temporal behavior of
empirical moments is not determined by the trait-evolution model alone. Even
under the same Brownian model, with the same variance parameter, the empirical
mean and empirical variance can display qualitatively different distributional
dynamics depending on the diversification history. This point is supported by
both the theoretical expressions and the simulations, and appears in both
settings considered here: fixed trees and random trees generated by
diversification models.

On a fixed tree, the effect of diversification history on the distributional
dynamics of the empirical moments of a Brownian trait is particularly clear.
As shown by Propositions~\ref{prop:piecewiseMean}
and~\ref{prop:piecewiseVar} and illustrated in
Figure~\ref{fig:simTreeDist}, both the variance of the empirical mean and the
expected empirical variance increase linearly between speciation and extinction
events, with trajectories punctuated by jumps at these events. In this
setting, trait stochasticity has a simple and regular effect on these
distributional summaries: it accounts for the linear increases between
diversification events and sets their scale through the Brownian variance
parameter. Diversification history, in contrast, drives both the timing and the
magnitude of the jumps.

The distributional behavior of the empirical mean and empirical variance on
random trees thus depends strongly on the underlying diversification process.
However, lineage homogeneity, a basic and common assumption in diversification
models, already constrains this dependence, notably by determining the expected
sign of the jumps at diversification events. It also
suffices to obtain general expressions for the quantities considered here in
the Brownian case. Conditionally on the number of extant lineages, the effect of
diversification on these quantities is entirely summarized by the expected MRCA
age of a uniformly sampled pair of distinct extant lineages.

In the generalized birth-death framework, the general identities obtained above
lead to explicit expressions for the variance of the empirical mean and the
expected empirical variance. This makes it possible to explore the effect of
various diversification scenarios on the dynamics of these quantities for a
Brownian trait. Even in the constant-rate case, the supercritical, critical and subcritical
regimes already produce contrasted dynamics: depending on the regime, these
quantities may show approximately linear growth, growth with decreasing slope,
convergence to a finite limit, or eventual decrease
(Figures~\ref{fig:meanVarianceConstRates}--\ref{fig:empiricalVarianceCond}). For the expected empirical variance, these contrasts
are confirmed by the asymptotic results
(Proposition~\ref{prop:asymp} and Corollary~\ref{cor:asympcond}). Figure~\ref{fig:variableRatesPaths} shows further that diversification
processes with variable rates can produce even more complex dynamics.

We considered the expected empirical variance both conditional on survival and
without conditioning, with the convention that an extinct tree has empirical
variance zero. Although the unconditional expectation has a less direct
biological interpretation, it shows how the expected disparity generated within
surviving trees interacts with the probability of extinction. The critical case
illustrates this interaction particularly well. Although the process becomes
extinct almost surely, the unconditional expected empirical variance does not
converge to zero, but to a positive finite limit. This reflects a balance
between the disparity accumulated within trees that survive up to a given time
and the decreasing probability of observing non-extinct trees.

In the constant birth-death case, Figures~\ref{fig:empiricalVarianceUncond} and~\ref{fig:empiricalVarianceCond}, together with the
asymptotic results of Proposition~\ref{prop:asymp} and Corollary~\ref{cor:asympcond}, show that the
dynamics of the expected empirical variance can change qualitatively depending
on whether or not this quantity is conditioned on survival. In the supercritical
regime, conditioning on survival does not change the qualitative behavior, and
the expected empirical variance grows approximately linearly in both cases. In
the critical regime, however, conditioning on survival changes the long-term
behavior from convergence to a finite limit to linear growth. In the
subcritical regime, it changes the behavior from eventual exponential decrease
to convergence to a positive finite limit.

Returning to the general lineage-homogeneous setting, a remarkable feature of the expressions of the variance of the empirical mean and of the expected empirical variance is that the expected MRCA age
contributes with opposite signs to the variance of the empirical mean
(Eq.~\ref{eq:meanVarianceCond}) and to the expected empirical variance
(Eq.~\ref{eq:varExpectCond}). Conditionally on the number of extant tips,
increasing the expected MRCA age therefore increases the variance of the
empirical mean but decreases the expected empirical variance, and conversely. After averaging 
over the number of tips, a related opposition remains visible, although the 
unconditioned expressions also involve the distribution of the number of extant tips.

Within the theoretical framework developed here, this negative association appears in the behavior of jumps at diversification events. Under the lineage-homogeneity assumption, the expected jumps in the variance of the empirical mean and in the expected empirical variance have opposite directions: they are nonnegative for the former and nonpositive for the latter.
 The same sign opposition is preserved in
the generalized birth-death case: although the final integral expressions are
more involved, both quantities decompose into a term proportional to time and a
term depending on the diversification process, and these two contributions
enter with opposite signs in the two expressions
(Propositions~\ref{prop:variance} and~\ref{prop:expect}).

This general behavior is illustrated by the figures showing the dynamics of the
variance of the empirical mean and of the expected empirical variance. In the
constant-rate birth-death examples conditional on survival
(Figures~\ref{fig:meanVarianceConstRates} and~\ref{fig:empiricalVarianceCond}),
the supercritical and subcritical cases show opposite behaviors for the two
quantities. Namely, in the supercritical regime, the variance of the empirical
mean increases with a decreasing slope, while the expected empirical variance
appears to grow linearly, whereas the opposite pattern is observed in the
subcritical regime. The critical regime is intermediate: both quantities are
close to linear, which does not contradict the negative association, since it may be expressed through
differences in slope.
For instance, in the variable-rate example
(Figure~\ref{fig:variableRatesPaths}), both the variance of the empirical mean
and the expected empirical variance are roughly piecewise linear, but intervals
with a higher slope for the variance of the empirical mean correspond to
intervals with a lower slope for the expected empirical variance, and
conversely.

One striking consequence of this sign opposition is that, intuitively,
diversification conditions that favor displacement of the empirical mean away
from the ancestral state tend to be associated with lower expected trait
disparity.
Such a relation between these two important features of trait evolution is
unexpected and, to our knowledge, has not been pointed out previously.

Whether this negative relationship is detectable in biological data remains an
open question. Combining fossil data with extant comparative data provides a
natural setting in which to address this issue. Fossils close to the origin of a
clade may help constrain the distribution of plausible early trait values,
while extant species provide direct information on the present empirical mean
and disparity of the clade. Comparing these two sources of information could
make it possible to assess whether large displacement from early trait values
is associated with reduced extant disparity. Such comparisons would, however,
require careful treatment of fossil sampling, phylogenetic placement, and
uncertainty in the initial state.

\section*{Declaration of Generative AI and AI-assisted technologies in the writing process}
During the preparation of this work the author used AI-assisted technologies in order to improve language and readability.
 After using this tool, the author reviewed
and edited the content as needed and takes full responsibility for the content
of the publication.

\begin{appendix}
\numberwithin{equation}{section}
\renewcommand{\theequation}{\thesection-\arabic{equation}}
\counterwithin{proposition}{section}
\renewcommand{\theproposition}{\thesection-\arabic{proposition}}

\section{Proof of Propositions~\ref{prop:piecewiseMean} and~\ref{prop:piecewiseVar}}
Between two diversification events (speciation or extinction), the set of extant lineages remains fixed and trait values evolve independently along each lineage. To distinguish configurations immediately before and after a diversification
event, we write, for any time $\ttot$, $\ttot^{-}$ and $\ttot^{+}$ for the instants
immediately before and immediately after $\ttot$, respectively.

Let $\ttot$ be a time, $\tdiv$ the time of the most recent diversification
event preceding $\ttot$ in $\tree$, and $n$ the number of lineages present on
$(\tdiv,\ttot)$. Assuming that $n\geq 1$, for each extant lineage $i$, we have
\[
\trait[\tree(\ttot)]^{(i)}=\trait[\tree(\tdiv^+)]^{(i)}+Y_i,
\]
where the random variables $Y_1,\ldots,Y_n$ are independent centered Gaussian
variables with variance $\sigma^2(\ttot-\tdiv)$, and are independent of the trait
values at time $\tdiv^+$, i.e., an instant after $\tdiv$. We denote $Y$ the vector $(Y_{i})_{1\leq i \leq n}$. Its empirical mean $\bar{Y}$ has variance $\frac{\sigma^2(\ttot-\tdiv)}{n}$ (Section~\ref{sec:ind}).

The variance of the empirical mean of the trait values at $\ttot$ is 
\[
\variance(\meanTrait[\tree(\ttot)]) = \variance(\meanTrait[\tree(\tdiv^+)]+\bar{Y}) = \variance(\meanTrait[\tree(\tdiv^+)])+\variance(\bar{Y}) = \variance(\meanTrait[\tree(\tdiv^+)])+\frac{\sigma^2(\ttot-\tdiv)}{n} 
\]
which proves Proposition~\ref{prop:piecewiseMean}.

Let us now assume that $n\geq 2$. By expansion of the empirical variance,
\[
\mavar{\tree(\ttot)} = \mavar{\tree(\tdiv^+)} + \mavar{Y}
+ \frac{2}{n-1}\sum_{i=1}^n
\left(\trait[\tree(\tdiv^+)]^{(i)}-\meanTrait[\tree(\tdiv^+)]\right)\left(Y_i-\bar Y\right).
\]
Since the increments $(Y_{i})_{1\leq i \leq n}$ are centered and independent both of the past and of one
another, the expectation of the third term is zero. Hence,
\[
\esp\left(\mavar{\tree(\ttot)}\mid \tree\right)
=
\esp\left(\mavar{\tree(\tdiv^+)}\mid \tree\right)
+
\esp\left(\mavar{Y}\right).
\]

where $\mavar{Y}$ denotes the empirical variance of $n$ independent centered Gaussian variables with variance $\sigma^2(\ttot-\tdiv)$ which is the case studied in Section~\ref{sec:ind}. We have
\[
\esp\left(\mavar{Y}\right)=\sigma^2(\ttot-\tdiv).
\]
Thus,
\[
\esp\left(
\mavar{\tree(\ttot)}\mid \tree
\right)
=
\esp\left(
\mavar{\tree(\tdiv^{+})}\mid \tree
\right)
+\sigma^2(\ttot-\tdiv),
\]
which proves Proposition~\ref{prop:piecewiseVar}.

\section{Expected jumps}\label{app:jumps}

At a speciation event (resp. at an extinction event), one of the extant trait values is duplicated (resp. removed). In both cases, this
induces an instantaneous change, that is, a jump, in the empirical mean and in the
empirical variance. The direction of the jump depends on the
value that is duplicated or removed. The following propositions describe their
expected behavior for lineage-homogeneous random trees.

\begin{proposition}\label{prop:jumpSpeMean}
Let $\treeRV$ be a lineage-homogeneous random tree. The jump at a
speciation time $\tspec$ in the conditional variance of the
empirical mean of a Brownian trait evolving along $\treeRV$ has nonnegative
expectation given that $\card{\tips[\treeRV(\tspecBef)]} \geq 1$; namely,
\[
\esp\left(
\variance\left(\meanTrait[\treeRV(\tspecAft)]\mid\treeRV\right)
-
\variance\left(\meanTrait[\treeRV(\tspecBef)]\mid\treeRV\right)
\mid
\card{\tips[\treeRV(\tspecBef)]} \geq 1
\right)
\geq 0.
\]
\end{proposition}

\begin{proposition}\label{prop:jumpExtMean}
Let $\treeRV$ be a lineage-homogeneous random tree. The jump at an extinction time 
$\texti$ in the conditional variance of the empirical mean of a Brownian trait evolving along 
$\treeRV$ has nonnegative expectation given that $\card{\tips[\treeRV(\textiBef)]} \geq 2$; 
 namely,
\[
\esp\left(
\variance\left(\meanTrait[\treeRV(\textiAft)]\mid\treeRV\right)
-
\variance\left(\meanTrait[\treeRV(\textiBef)]\mid\treeRV\right)
\mid
\card{\tips[\treeRV(\textiBef)]} \geq 2
\right)
\geq 0.
\]
\end{proposition}

\begin{proposition}\label{prop:jumpSpeVar}
Let $\treeRV$ be a lineage-homogeneous random tree. The jump at a
speciation time $\tspec$ in the conditional expectation of the empirical variance of a Brownian trait evolving along 
$\treeRV$ has nonpositive expectation given that $\card{\tips[\treeRV(\tspecBef)]} \geq 2$; 
 namely,
\[
\esp\left(
\esp\left(\mavar{\treeRV(\tspecAft)}\mid \treeRV\right)
-
\esp\left(\mavar{\treeRV(\tspecBef)}\mid \treeRV\right)\mid \card{\tips[\treeRV(\tspecBef)]} \geq 2
\right)
\leq 0.
\]
\end{proposition}

In the case where $\card{\tips[\treeRV(\tspecBef)]} = 1$, since by convention
the empirical variance of a single value is zero and the speciation event
results in two equal values, the jump expectation is zero.

\begin{proposition}\label{prop:jumpExtVar}
Let $\treeRV$ be a lineage-homogeneous random tree. The jump at  an extinction time 
$\texti$ in the conditional expectation of the empirical variance of a Brownian trait evolving along 
$\treeRV$ has zero expectation given that $\card{\tips[\treeRV(\textiBef)]} \geq 3$; 
 namely,
\[
\esp\left(
\esp\left(\mavar{\treeRV(\textiAft)}\mid \treeRV\right)
-
\esp\left(\mavar{\treeRV(\textiBef)}\mid \treeRV\right)\mid \card{\tips[\treeRV(\textiBef)]} \geq 3
\right)
= 0.
\]
\end{proposition}

If $\card{\tips[\treeRV(\textiBef)]}\leq 2$, then $\card{\tips[\treeRV(\textiAft)]}\leq 1$ and, by convention, the empirical variance is zero after the extinction.
If $\card{\tips[\treeRV(\textiBef)]}=2$, the empirical variance is nonnegative before the event, so the expected jump is nonpositive. If
$\card{\tips[\treeRV(\textiBef)]}=1$, the empirical variance is zero both
before and after the event, so the expected jump is zero.

Together, these four propositions account for the qualitative behavior of the
jumps observed in Figure~\ref{fig:simTreeDist}.

\subsection{Proofs of Proposition~\ref{prop:jumpSpeMean} and~\ref{prop:jumpSpeVar}}
Let $\tspec$ be a speciation time and let us assume that the tips of $\treeRV(\tspecBef)$ are labelled
$1,\ldots,n$, that these tips keep the same label in $\treeRV(\tspecAft)$ and that the label of the new tip arising from the speciation event is $n+1$. 

Let $L$ be the random label of the tip of
$\treeRV(\tspecBef)$ on which the speciation event occurs. By lineage homogeneity, the distribution of $L$ is uniform over $\{1,\ldots, n\}$ conditionally on $\card{\tips[\treeRV(\tspecBef)]}=n$.

The MRCA ages of $\treeRV(\tspecAft)$ are obtained from those of 
$\treeRV(\tspecBef)$ by
\begin{itemize}
\item $\mrcaAgeVa[\treeRV(\tspecAft)] (i,j) = \mrcaAgeVa[\treeRV(\tspecBef)] (i,j)$ for all tips $i$ and $j$ in $\tips[\treeRV(\tspecBef)]$,
\item $\mrcaAgeVa[\treeRV(\tspecAft)] (n+1,j) = \mrcaAgeVa[\treeRV(\tspecAft)] (j, n+1) = \mrcaAgeVa[\treeRV(\tspecBef)] ( L, j)$ for all tips $j$ in $\tips[\treeRV(\tspecBef)]$, in particular $\mrcaAgeVa[\treeRV(\tspecAft)] (n+1, L) = \mrcaAgeVa[\treeRV(\tspecAft)] ( L, n+1)=\tspec$,
\item $\mrcaAgeVa[\treeRV(\tspecAft)](n+1,n+1)=\tspec$.
\end{itemize}

From Section~\ref{sec:fixed:phylo:mean}, if $n\geq 1$, the difference in the variance of the empirical mean after and before the
speciation is
\begin{align*}
\variance(\meanTrait[\treeRV(\tspecAft)]\mid\treeRV)-\variance(\meanTrait[\treeRV(\tspecBef)]\mid\treeRV)
&=
\frac{\sigma^2}{(n+1)^2}
\sum_{1\leq i,j\leq n+1}
\mrcaAgeVa[\treeRV(\tspecAft)](i,j)
-
\frac{\sigma^2}{n^2}
\sum_{1\leq i,j\leq n}
\mrcaAgeVa[\treeRV(\tspecBef)](i,j)
\\*&=
\frac{\sigma^2}{n^{2}(n+1)^2}
\Bigg(
n^{2}
\bigg(
\sum_{1\leq i,j\leq n}
\mrcaAgeVa[\treeRV(\tspecBef)](i,j)
+
2\sum_{i=1}^{n}
\mrcaAgeVa[\treeRV(\tspecBef)](i,L)
+
\tspec
\bigg)\\*
&-
(n+1)^2
\sum_{1\leq i,j\leq n}
\mrcaAgeVa[\treeRV(\tspecBef)](i,j)
\Bigg)
\\*&=
\frac{\sigma^2}{n^{2}(n+1)^2}
\left(
-(2n+1)
\sum_{1\leq i, j\leq n}
\mrcaAgeVa[\treeRV(\tspecBef)](i,j)
+
2n^{2}
\sum_{i=1}^{n}
\mrcaAgeVa[\treeRV(\tspecBef)](i,L)+
n^{2}\tspec
\right)
\\*&=
\frac{\sigma^2}{n^{2}(n+1)^2}
\Bigg(
n(n-1)\tspec
-(2n+1)
\sum_{1\leq i\neq j\leq n}
\mrcaAgeVa[\treeRV(\tspecBef)](i,j)
\\*&+
2n^{2}
\sum_{i=1, i\neq L}^{n}
\mrcaAgeVa[\treeRV(\tspecBef)](i,L)
\Bigg).
\end{align*}
The last line comes from the fact that $\mrcaAgeVa[\treeRV(\tspecBef)](j,j)=\tspec$ for every $1\leq j\leq n$.

The exchangeability property stated at the beginning of
Section~\ref{sec:linHom}, together with the fact that $L$ is uniformly chosen
among the $n$ extant lineages, implies that, for all tips $i\neq L$,
\[
\esp\left(
\mrcaAgeVa[\treeRV(\tspecBef)](i,L)
\mid
\card{\tips[\treeRV(\tspecBef)]}= n
\right)
=
\esp(\mrcaAgeRVU[\treeRV(\tspecBef)]\mid \card{\tips[\treeRV(\tspecBef)]}=n).
\]

From Equation~\ref{eq:varMean} and the identity above, we get
\begin{multline*}
\esp\left(
\variance\left(\meanTrait[\treeRV(\tspecAft)]\mid\treeRV\right)
-
\variance\left(\meanTrait[\treeRV(\tspecBef)]\mid\treeRV\right)
\mid
\card{\tips[\treeRV(\tspecBef)]} =n
\right)
\\=
\frac{\sigma^2}{n^{2}(n+1)^2}
\left(
n(n-1)\left(\tspec
-
\esp(\mrcaAgeRVU[\treeRV(\tspecBef)]\mid \card{\tips[\treeRV(\tspecBef)]}=n)
\right)
\right)
{\geq} 0.
\end{multline*}
This proves Proposition~\ref{prop:jumpSpeMean}, since the argument holds for
every $n\geq 1$.

In order to study the jump in expected empirical variance, let us assume that $\card{\tips[\treeRV(\tspecBef)]} = n \geq 2$. Proposition~\ref{prop:exp_cov} yields
\begin{multline*}
\esp\left(\mavar{\treeRV(\tspecAft)}\mid \treeRV\right)-\esp\left(\mavar{\treeRV(\tspecBef)}\mid \treeRV\right) 
= \sigma^{2}\left(\tspec-\frac{2\sum_{1\leq i<j\leq n+1}\mrcaAgeVa[\treeRV(\tspecAft)](i,j)}{(n+1)n}\right)-\sigma^{2}\left(\tspec-\frac{2\sum_{1\leq i<j\leq n}\mrcaAgeVa[\treeRV(\tspecBef)](i,j)}{n(n-1)}\right)
\\=\frac{2\sigma^{2}}{(n-1)n(n+1)}\left(2\sum_{1\leq i<j\leq n}\mrcaAgeVa[\treeRV(\tspecBef)](i,j)-(n-1)\sum_{1\leq i\leq n, i\neq L}\mrcaAgeVa[\treeRV(\tspecBef)](i, L) -(n-1)\tspec\right).
\end{multline*}
Taking conditional expectation of the difference above using the same considerations as in the case of the variance of the empirical mean, leads to
\begin{multline*}
\esp\left(
\esp\left(\mavar{\treeRV(\tspecAft)}{\mid} \treeRV\right)
-
\esp\left(\mavar{\treeRV(\tspecBef)}{\mid} \treeRV\right){\mid} \card{\tips[\treeRV(\tspecBef)]} = n
\right) 
=\frac{2\sigma^{2}}{(n-1)n(n+1)}\Bigg(2\sum_{1\leq i<j\leq n}\esp(\mrcaAgeVa[\treeRV(\tspecBef)](i,j)\mid \card{\tips[\treeRV(\tspecBef)]} = n)\\*
-(n-1)\sum_{i=1, i\neq L}^{n}\esp(\mrcaAgeVa[\treeRV(\tspecBef)](i,L)\mid \card{\tips[\treeRV(\tspecBef)]} = n) -(n-1)\tspec\Bigg)
\\*=\frac{2\sigma^{2}}{n(n+1)}\left(\esp(\mrcaAgeRVU[\treeRV(\tspecBef)]\mid \card{\tips[\treeRV(\tspecBef)]}=n)-\tspec\right) {\leq} 0.
\end{multline*}
Since this holds for every $n\geq 2$, Proposition~\ref{prop:jumpSpeVar} follows.

\subsection{Proofs of Proposition~\ref{prop:jumpExtMean} and~\ref{prop:jumpExtVar} }
Let $\texti$ be an extinction time and let us assume that the tips of $\treeRV(\textiBef)$ are labelled
$1,\ldots,n$, and let $L$ be the random label of the tip of
$\treeRV(\textiBef)$ on which the extinction event occurs. We assume here that the remaining tips of $\treeRV(\textiAft)$ keep the label they had in $\treeRV(\textiBef)$.

The MRCA ages of 
$\treeRV(\textiAft)$ are then obtained from those of 
$\treeRV(\textiBef)$ by $\mrcaAgeVa[\treeRV(\textiAft)] (i,j) = \mrcaAgeVa[\treeRV(\textiBef)] (i,j)$ for all tips $i$ and $j$ in $\{1,\ldots,n\}\setminus\{L\}$.

If $n\geq 2$, the difference in the variance of the empirical mean after and before the
extinction is
\begin{align*}
\variance(\meanTrait[\treeRV(\textiAft)]\mid\treeRV)-\variance(\meanTrait[\treeRV(\textiBef)]\mid\treeRV)
&=
\frac{\sigma^2}{(n-1)^2}
\sum_{j=1,j\neq L}^{n}\sum_{i=1,i\neq L}^{n}
\mrcaAgeVa[\treeRV(\textiAft)](i,j)
-
\frac{\sigma^2}{n^2}
\sum_{j=1}^{n}\sum_{i=1}^{n}
\mrcaAgeVa[\treeRV(\textiBef)](i,j)
\\*&=
\frac{\sigma^2}{n^{2}(n-1)^2}
\Bigg(
n^{2}
\sum_{j=1}^{n}\sum_{i=1}^{n}
\mrcaAgeVa[\treeRV(\textiBef)](i,j)
-
2n^{2}
\sum_{i=1}^{n}
\mrcaAgeVa[\treeRV(\textiBef)](i,L)
+n^{2}\texti
\\*
&\quad-(n-1)^2
\sum_{j=1}^{n}\sum_{i=1}^{n}
\mrcaAgeVa[\treeRV(\textiBef)](i,j)
\Bigg)
\\&=
\frac{\sigma^2}{n^{2}(n-1)^2}
\Bigg(
(2n-1)
\sum_{j=1}^{n}\sum_{i=1, i\neq j}^{n}
\mrcaAgeVa[\treeRV(\textiBef)](i,j)
\\*&\quad-
2n^{2}
\sum_{i=1, i\neq L}^{n}
\mrcaAgeVa[\treeRV(\textiBef)](i,L)
+n(n-1)\texti\Bigg)
\end{align*}
From Equation~\ref{eq:varMean} and the identity above, we get
\begin{flalign*}
&
\esp\left(
\variance\left(\meanTrait[\treeRV(\textiAft)]\mid\treeRV\right)
-
\variance\left(\meanTrait[\treeRV(\textiBef)]\mid\treeRV\right)
\mid
\card{\tips[\treeRV(\textiBef)]} =n
\right)
\\*
& =
\frac{\sigma^2}{n^{2}(n-1)^2}
\Bigg(
(2n-1)
\sum_{j=1}^{n}\sum_{i=1,i\neq j}^{n}
\esp\left(\mrcaAgeVa[\treeRV(\textiBef)](i,j)\mid
\card{\tips[\treeRV(\textiBef)]}=n
\right)
\\*
&\quad -
2n^2
\sum_{i=1,i\neq L}^{n}
\esp\left(\mrcaAgeVa[\treeRV(\textiBef)](i,L)\mid
\card{\tips[\treeRV(\textiBef)]}=n
\right)
+n(n-1)\texti
\Bigg)
\\*
& =
\frac{\sigma^2}{n(n-1)}
\left(\texti-\esp(\mrcaAgeRVU[\treeRV(\textiBef)]\mid \card{\tips[\treeRV(\textiBef)]}=n)\right)
\geq 0,
\end{flalign*}
which holds for all $n\geq 2$ and proves Proposition~\ref{prop:jumpExtMean}.

In order to prove Proposition~\ref{prop:jumpExtVar}, let us assume that $\card{\tips[\treeRV(\textiBef)]} = n \geq 3$. Proposition~\ref{prop:exp_cov} yields
\begin{align*}
\esp\left(\mavar{\treeRV(\textiAft)}\mid \treeRV\right)
-
\esp\left(\mavar{\treeRV(\textiBef)}\mid \treeRV\right) &= \sigma^{2}\left(\texti-2\frac{\sum_{1\leq i<j\leq n, i, j\neq L}\mrcaAgeVa[\treeRV(\textiAft)](i,j)}{(n-2)(n-1)}\right)-\sigma^{2}\left(\texti-2\frac{\sum_{1\leq i<j\leq n}\mrcaAgeVa[\treeRV(\textiBef)](i,j)}{n(n-1)}\right)
\\&=\frac{2\sigma^{2}}{(n-2)(n-1)n}\left(-n\sum_{1\leq i<j\leq n, i, j\neq L}\mrcaAgeVa[\treeRV(\textiAft)](i,j)+(n-2)\sum_{1\leq i<j\leq n}\mrcaAgeVa[\treeRV(\textiBef)](i,j)\right)
\\&=\frac{2\sigma^{2}}{(n-2)(n-1)n}\left(n\sum_{i=1, i\neq L}^{n}\mrcaAgeVa[\treeRV(\textiBef)](i,L)-\sum_{j=1}^{n}\sum_{i=1,i\neq j}^{n}\mrcaAgeVa[\treeRV(\textiBef)](i,j)\right)
\end{align*}

Taking conditional expectation of the difference above, and using the same argument as for the variance of the empirical mean case, we get for all $n\geq 3$,
{\begin{multline*}
\esp\left(
\esp\left(\mavar{\treeRV(\textiAft)}\mid \treeRV\right)
-
\esp\left(\mavar{\treeRV(\textiBef)}\mid \treeRV\right)\mid \card{\tips[\treeRV(\textiBef)]} = n
\right) 
\\*= \frac{2\sigma^{2}}{(n-2)(n-1)n}\left(n\sum_{i=1, i\neq L}^{n}\esp(\mrcaAgeVa[\treeRV(\textiBef)](i,L)\mid \card{\tips[\treeRV(\textiBef)]} = n)-\sum_{j=1}^{n}\sum_{i=1,i\neq j}^{n}\esp(\mrcaAgeVa[\treeRV(\textiBef)](i,j)\mid \card{\tips[\treeRV(\textiBef)]} = n)\right) = 0
\end{multline*}}
Since this holds for every $n\geq 3$, Proposition~\ref{prop:jumpExtVar} follows.

\section{Proof of Proposition~\ref{prop:coal}}
This argument is a straightforward adaptation of the proof of Theorem~4.1 in \citet{Stadler2009}, which gives the probability that two random lineages coalesce at the $k$-th speciation event or, equivalently, that their MRCA corresponds to the $k$-th speciation event.

Let $\tfir\le \ttot$, and assume that there are $m$ and $n$ lineages present at times $\tfir$ and $\ttot$, respectively, with $1\leq m\le n$ and $n\geq 2$. 
Ordering the speciation events from the start of the process, the MRCA age of two extant lineages at $\ttot$ is at most $\tfir$ if and only if they do not coalesce at any of the $m$-th,\ldots, $(n-1)$-th speciation events.
Under the homogeneity assumption, two random lineages at time $\ttot$ do not coalesce at the most recent speciation event before $\ttot$, i.e., the $(n-1)$-th, with probability $1-\frac{1}{\binom{n}{2}}$. In this case, we are left with $n-1$ lineages, two of which correspond to the sampled lineages. Iterating this argument yields
\[
\prob\left(\mrcaAgeRVU[\treeRV(\ttot)]\leq \tfir \mid \effRec[\treeRV(\ttot)](\tfir) = m , \effRec[\treeRV(\ttot)](\ttot) = n\right)
=
\prod_{\ell=m+1}^{n}\left(1-\frac{1}{\binom{\ell}{2}}\right)
=
\prod_{\ell=m+1}^{n}\left(1-\frac{2}{\ell(\ell-1)}\right).
\]
Since
\[
1-\frac{2}{\ell(\ell-1)}=\frac{\ell(\ell-1)-2}{\ell(\ell-1)}
=\frac{(\ell+1)(\ell-2)}{\ell(\ell-1)},
\]
the product telescopes:
\begin{align*}
\prob\left(\mrcaAgeRVU[\treeRV(\ttot)]\leq \tfir \mid \effRec[\treeRV(\ttot)](\tfir) = m , \effRec[\treeRV(\ttot)](\ttot) = n\right)
&=\prod_{\ell=m+1}^{n}\frac{(\ell+1)(\ell-2)}{\ell(\ell-1)}
\\&=
\frac{\prod_{\ell=m+1}^{n}(\ell+1)}{\prod_{\ell=m+1}^{n}\ell}\times
\frac{\prod_{\ell=m+1}^{n}(\ell-2)}{\prod_{\ell=m+1}^{n}(\ell-1)}
\\&=
\frac{n+1}{m+1}\times\frac{m-1}{n-1}
=\left(1+\frac{2}{n-1}\right)\left(1-\frac{2}{m+1}\right).
\end{align*}

\section{Proof of Proposition~\ref{prop:mrcaDistCond}}\label{app:mrcaDistCond}
By setting
\[
y=1-\mexp{-(\ftime(\ttot)-\ftime(\tfir))}\;\mbox{ and }\;
x=\frac{\mexp{-(\ftime(\ttot)-\ftime(\tfir))}(1-\mexp{-\ftime(\tfir)})}{y},
\]
Equation~\ref{eq:trans} becomes
\[
\prob_{\ttot}\big(\effRec[\treeRV(\ttot)](\tfir)=m,\ \effRec[\treeRV(\ttot)](\ttot)=n\big)
=
\frac{\mexp{-\ftime(\tfir)}}{1-\mexp{-\ftime(\tfir)}}\,
y^n\binom{n-1}{m-1}x^m.
\]

Let us compute
\[
\prob_{\ttot}\left(\mrcaAgeRVU[\treeRV(\ttot)]\le \tfir,\ \effRec[\treeRV(\ttot)](\ttot)=n\right)
=
\sum_{m=1}^{n}
\prob_{\ttot}\left(\mrcaAgeRVU[\treeRV(\ttot)]\le \tfir \mid \effRec[\treeRV(\ttot)](\tfir)=m,\effRec[\treeRV(\ttot)](\ttot)=n\right)
\prob_{\ttot}\big(\effRec[\treeRV(\ttot)](\tfir)=m,\effRec[\treeRV(\ttot)](\ttot)=n\big).
\]
Proposition~\ref{prop:coal} gives us that
\[
\prob_{\ttot}\left(\mrcaAgeRVU[\treeRV(\ttot)]\le \tfir \mid \effRec[\treeRV(\ttot)](\tfir)=m,\effRec[\treeRV(\ttot)](\ttot)=n\right)
=
\left(1+\frac{2}{n-1}\right)\left(1-\frac{2}{m+1}\right).
\]
Therefore
\[
\prob_{\ttot}\left(\mrcaAgeRVU[\treeRV(\ttot)]\le \tfir,\ \effRec[\treeRV(\ttot)](\ttot)=n\right)
=
\frac{\mexp{-\ftime(\tfir)}}{1-\mexp{-\ftime(\tfir)}}
y^n
\left(1+\frac{2}{n-1}\right)
S_{n}(x),
\]
where
\[
S_{n}(x)
=
\sum_{m=1}^{n}
\left(1-\frac{2}{m+1}\right)
\binom{n-1}{m-1}x^m.
\]
Using
\[
\sum_{m=1}^{n}\binom{n-1}{m-1}x^m=x(1+x)^{n-1},
\]
and
\[
\sum_{m=1}^{n}\binom{n-1}{m-1}\frac{x^m}{m+1}
=
\int_0^1 xu(1+xu)^{n-1}\mder u
=
\frac{(1+x)^n(nx-1)+1}{xn(n+1)},
\]
we obtain
\begin{align}
S_n(x)
&=x(1+x)^{n-1}-2\int_0^1 xu(1+xu)^{n-1}\mder u\label{eq:SnA}\\ 
&= 
\frac{1}{xn(n+1)}
\left((1+x)^{n-1}
\left(x^{2}n(n-1)-2\left(x(n-1)-1\right)\right)-2\right).\label{eq:SnB}
\end{align}

Substituting $S_{n}(x)$ using Equation~\ref{eq:SnB} yields
\begin{align*}
\prob_{\ttot}\left(\mrcaAgeRVU[\treeRV(\ttot)]\le \tfir,\ \effRec[\treeRV(\ttot)](\ttot)=n\right)
&=\frac{\mexp{-\ftime(\tfir)}}{1-\mexp{-\ftime(\tfir)}}
y^n
\left(1+\frac{2}{n-1}\right)
\frac{1}{xn(n+1)}\\&\quad\times\left((1+x)^{n-1}(x^{2}n(n-1)-2\left(x(n-1)-1\right))-2\right)\\
&=\frac{\mexp{-\ftime(\tfir)}y(y(1+x))^{n-1}}{1-\mexp{-\ftime(\tfir)}}
\left(x-\frac{2}{n}\left(1-\frac{1}{(n-1)x}\left(1-\frac{1}{(1+x)^{n-1}}\right)\right)\right).
\end{align*}

Substituting back $x$ and $y$, noting that $y(1+x) =1-\mexp{-\ftime(\ttot)}$, and dividing by $\prob_{\ttot}\left(\effRec[\treeRV(\ttot)](\ttot)=n\right) = \mexp{-\ftime(\ttot)}\left(1-\mexp{-\ftime(\ttot)}\right)^{n-1}$ yields after simplification

\[ \prob_{\ttot}\left( \mrcaAgeRVU[\treeRV(\ttot)]\le \tfir \mid \effRec[\treeRV(\ttot)](\ttot)=n \right) 
 = 1 - \frac{ 2\left(\mexp{\ftime(\ttot)}-\mexp{\ftime(\tfir)}\right) }{ n\left(\mexp{\ftime(\tfir)}-1\right) }\left(1 - \frac{ \left(\mexp{\ftime(\ttot)}-\mexp{\ftime(\tfir)}\right)}{ (n-1)\left(\mexp{\ftime(\tfir)}-1\right) } \left[ 1- \left( \frac{ \mexp{\ftime(\ttot)}-\mexp{\ftime(\tfir)} }{ \mexp{\ftime(\ttot)}-1 } \right)^{n-1} \right]\right).
\]
This proves the proposition.

\section{Proof of Proposition~\ref{prop:mrcaDist}}
We shall first compute
\[
\prob_{\ttot}\left(\mrcaAgeRVU[\treeRV(\ttot)]\le \tfir,\ \effRec[\treeRV(\ttot)](\ttot)\ge 2\right)
=
\sum_{n=2}^{\infty}\sum_{m=1}^{n}
\prob_{\ttot}\left(\mrcaAgeRVU[\treeRV(\ttot)]\le \tfir \mid \effRec[\treeRV(\ttot)](\tfir)=m,\effRec[\treeRV(\ttot)](\ttot)=n\right)
\prob_{\ttot}\big(\effRec[\treeRV(\ttot)](\tfir)=m,\effRec[\treeRV(\ttot)](\ttot)=n\big).
\]
Under the same notations as in Section~\ref{app:mrcaDistCond}, we have
\[
\prob_{\ttot}\left(\mrcaAgeRVU[\treeRV(\ttot)]\le \tfir,\ \effRec[\treeRV(\ttot)](\ttot)\ge 2\right)
=
\frac{\mexp{-\ftime(\tfir)}}{1-\mexp{-\ftime(\tfir)}}
\sum_{n=2}^{\infty}
y^n
\left(1+\frac{2}{n-1}\right)
S_{n}(x).
\]

Substituting $S_{n}(x)$ using Equation~\ref{eq:SnA} yields
\begin{multline*}
\prob_{\ttot}\left(\mrcaAgeRVU[\treeRV(\ttot)]\le \tfir,\ \effRec[\treeRV(\ttot)](\ttot)\ge 2\right)\\
=
\frac{\mexp{-\ftime(\tfir)}}{1-\mexp{-\ftime(\tfir)}}
\left(
xy\sum_{n=2}^{\infty}
\left(1+\frac{2}{n-1}\right)\big(y(1+x)\big)^{n-1}
-
2\int_0^1 xuy
\sum_{n=2}^{\infty}
\left(1+\frac{2}{n-1}\right)\big(y(1+xu)\big)^{n-1}\mder u
\right).
\end{multline*}

Let us remark that since
\[0\leq y(1+xu)\leq y(1+x)=1-\mexp{-\ftime(\ttot)}<1\mbox{ for all } u\in[0,1],\]
the identity
\[
\sum_{n=2}^{\infty}\left(1+\frac{2}{n-1}\right)a^{n-1}
=
\frac{a}{1-a}-2\ln(1-a),
\] 
holds for $a=y(1+x)$ and $a=y(1+xu)$ and in the latter case, the corresponding series converges uniformly on $u\in[0,1]$, so it may be integrated termwise.

We get
\begin{multline*}
\prob_{\ttot}\left(\mrcaAgeRVU[\treeRV(\ttot)]\le \tfir,\ \effRec[\treeRV(\ttot)](\ttot)\ge 2\right)
=
\frac{\mexp{-\ftime(\tfir)}}{1-\mexp{-\ftime(\tfir)}}
\Bigg[
xy\left(
\frac{y(1+x)}{1-y(1+x)}
-2\ln\big(1-y(1+x)\big)
\right)\\*
-2\int_0^1 xuy
\left(
\frac{y(1+xu)}{1-y(1+xu)}
-2\ln\big(1-y(1+xu)\big)
\right)\mder u
\Bigg].
\end{multline*}

Since
\begin{equation*}
\frac{\mexp{-\ftime(\tfir)}}{1-\mexp{-\ftime(\tfir)}}\,xy
=
\mexp{-\ftime(\ttot)},
\;
\frac{\mexp{-\ftime(\tfir)}}{1-\mexp{-\ftime(\tfir)}}\,xuy
=
u \mexp{-\ftime(\ttot)},
y(1+x)=1-\mexp{-\ftime(\ttot)},
\;
1-y(1+xu)=\mexp{-\ftime(\ttot)}g(u),\\
\end{equation*}
with $g(u)=\mexp{\ftime(\tfir)}-u\left(\mexp{\ftime(\tfir)}-1\right)$,
we get
\begin{multline*}
\prob_{\ttot}\left(\mrcaAgeRVU[\treeRV(\ttot)]\le \tfir,\ \effRec[\treeRV(\ttot)](\ttot)\ge 2\right)
=\\
\mexp{-\ftime(\ttot)}
\left(
\frac{1-\mexp{-\ftime(\ttot)}}{\mexp{-\ftime(\ttot)}}
-2\ln\left(\mexp{-\ftime(\ttot)}\right)
\right)
-2\int_0^1
u
\left(
\frac{1}{g(u)}-\mexp{-\ftime(\ttot)}
\right)\mder u
+
4\mexp{-\ftime(\ttot)}
\int_0^1
u\left(-\ftime(\ttot)+\ln\left(g(u)\right)\right)\mder u.
\end{multline*}

Since $\int_0^1 u\mder u=\frac12$, this reduces to
\[
\prob_{\ttot}\left(\mrcaAgeRVU[\treeRV(\ttot)]\le \tfir,\ \effRec[\treeRV(\ttot)](\ttot)\ge 2\right)
=
1
-
2\int_0^1
\frac{u\mder u}{g(u)}
+
4\mexp{-\ftime(\ttot)}
\int_0^1
u\ln\left(g(u)\right)\mder u.
\]

It remains to evaluate the two elementary integrals
\[
\int_0^1
\frac{u\mder u}{g(u)}=\int_0^1
\frac{u\mder u}{\mexp{\ftime(\tfir)}-u(\mexp{\ftime(\tfir)}-1)}
=
\frac{\ftime(\tfir)\mexp{\ftime(\tfir)}}{(\mexp{\ftime(\tfir)}-1)^2}
-
\frac{1}{\mexp{\ftime(\tfir)}-1},
\]
and
\[
\int_0^1
u\ln\left(g(u)\right)\mder u = \int_0^1 u\ln\big(\mexp{\ftime(\tfir)}-u(\mexp{\ftime(\tfir)}-1)\big)\mder u
=
\frac{-1+4\mexp{\ftime(\tfir)}-3\mexp{2\ftime(\tfir)}+2\ftime(\tfir)\mexp{2\ftime(\tfir)}}{4(\mexp{\ftime(\tfir)}-1)^2},
\]
 and to divide by $\prob_{\ttot}\left(\effRec[\treeRV(\ttot)](\ttot)\ge 2\right) = 1-\mexp{-\ftime(\ttot)}$ to obtain the final expression
\[
\prob_{\ttot}\left(\mrcaAgeRVU[\treeRV(\ttot)]\le \tfir \mid \ \effRec[\treeRV(\ttot)](\ttot)\ge 2\right)
=
1
-
2\frac{\left(1-\mexp{-(\ftime(\ttot)-\ftime(\tfir))}\right)
\left(1-\mexp{\ftime(\tfir)}+\ftime(\tfir)\mexp{\ftime(\tfir)}\right)}
{(1-\mexp{-\ftime(\ttot)})(\mexp{\ftime(\tfir)}-1)^2}.
\]

\section{Some facts about the dilogarithm}\label{app:dilog}

We provide here some basic properties of the dilogarithm function. For further information, we refer the reader to \citet{Morris1979, Lewin1982, Zagier2007}. 

We consider here the real-valued dilogarithm. It may be defined for
$z\leq 1$ by
\begin{equation}\label{eq:dilog:int}
\operatorname{Li}_2(z)
=
-\int_0^z \frac{\log(1-u)}{u}\mder u,
\end{equation}
where the integral is improper but finite at $z=1$. For
$|z|\leq 1$,
\begin{equation}\label{eq:dilog:ser}
\operatorname{Li}_2(z)
=
\sum_{k=1}^{\infty}\frac{z^k}{k^2}.
\end{equation}
In particular,
\begin{equation}\label{eq:dilog:vun}
\operatorname{Li}_2(1)=\frac{\pi^2}{6}.
\end{equation}
For real $z>1$, the above integral is no longer real-valued without choosing a
complex branch of the logarithm. This issue does not arise in the present
paper, since all arguments of $\operatorname{Li}_2$ that occur below belong to
$(-\infty,1]$.

For $z<1$, $z\neq 0$, differentiation under the integral sign gives
\begin{equation*}
\frac{d}{dz}\operatorname{Li}_2(z)
=
-\frac{\log(1-z)}{z}.
\end{equation*}

The dilogarithm function satisfies the following functional equations, referred to as \emph{reflection properties} in \citet{Zagier2007}:
\begin{align}
\operatorname{Li}_2\left(\frac{1}{x}\right)
&= -\frac{\pi^{2}}{6}-\frac{\log^{2}(-x)}{2}-\operatorname{Li}_2(x),
\quad \text{for } x<0, \label{eq:dilogrefA}\\
\operatorname{Li}_2(1-x)
&= \frac{\pi^{2}}{6}-\log(x)\log(1-x)-\operatorname{Li}_2(x),
\quad \text{for } 0<x<1.\label{eq:dilogrefB}
\end{align}

Using Equation~\ref{eq:dilogrefB}, together with the series expansions of $\log(1-x)$ and $\operatorname{Li}_2(x)$, yields, as $x\to0^+$,
\begin{equation}\label{eq:dilog:dev}
\operatorname{Li}_2(1-x)
= \frac{\pi^{2}}{6}+x\log(x)-x+\frac{x^{2}\log(x)}{2}-\frac{x^{2}}{4}+o(x^{2}).
\end{equation}

\section{Proof of Proposition~\ref{prop:variance}}\label{app:variance}

Using Equation~\ref{eq:varMean} to sum over the possible numbers of extant
tips, we get

\begin{align*}
\variance(\meanTrait[\treeRV(\ttot)]\mid \card{\tips[\treeRV(\ttot)]}\geq 1)
 &= {\sum_{n=1}^{\infty}\variance(\meanTrait[\treeRV(\ttot)]\mid \card{\tips[\treeRV(\ttot)]}=n)\prob_{\ttot}(\effRec[\treeRV(\ttot)](\ttot)=n)}\\
 &={\sigma^{2}\ttot\prob_{\ttot}(\effRec[\treeRV(\ttot)](\ttot)=1)+\sum_{n=2}^{\infty}\frac{\sigma^2}{n}\left(\ttot+(n-1)\esp(\mrcaAgeRVU[\treeRV(\ttot)]\mid \card{\tips[\treeRV(\ttot)]}=n)\right)\prob_{\ttot}(\effRec[\treeRV(\ttot)](\ttot)=n)}\\
 &={\sigma^{2}\ttot\prob_{\ttot}(\effRec[\treeRV(\ttot)](\ttot)=1)+\sum_{n=2}^{\infty}\frac{\sigma^2}{n}\left(\ttot+(n-1)\int_{0}^{\ttot} (1-\distri{\mrcaAgeRVU[\treeRV(\ttot)]\mid_{\overset{=}{n}}}(\tfir))\mder \tfir\right)\prob_{\ttot}(\effRec[\treeRV(\ttot)](\ttot)=n)}\\
&=\sigma^{2}\left(\ttot-\sum_{n=2}^{\infty}\frac{n-1}{n}\prob_{\ttot}(\effRec[\treeRV(\ttot)](\ttot)=n)\int_{0}^{\ttot}\distri{\mrcaAgeRVU[\treeRV(\ttot)]\mid_{\overset{=}{n}}}(\tfir)\mder \tfir\right)\\
 \end{align*}

Under the notation of Appendix~\ref{app:mrcaDistCond}, we have that for all $n\geq 2$,
\[
\prob_{\ttot}(\effRec[\treeRV(\ttot)](\ttot)=n)\distri{\mrcaAgeRVU[\treeRV(\ttot)]\mid_{\overset{=}{n}}}(\tfir) = \prob_{\ttot}\left(\mrcaAgeRVU[\treeRV(\ttot)]\leq \tfir, \effRec[\treeRV(\ttot)](\ttot)= n\right) = 
\frac{\mexp{-\ftime(\tfir)}}{1-\mexp{-\ftime(\tfir)}}y^n\frac{n+1}{n-1}S_{n}(x).
\]

Since the summands in the left-hand side are non-negative, Tonelli's theorem
allows us to exchange the sum over $n$ and the integral over $\tfir$. It follows that
\begin{align*}
\sum_{n=2}^{\infty}\frac{n-1}{n}\prob_{\ttot}(\effRec[\treeRV(\ttot)](\ttot)=n)\int_{0}^{\ttot}\distri{\mrcaAgeRVU[\treeRV(\ttot)]\mid_{\overset{=}{n}}}(\tfir)\mder \tfir
&=\sum_{n=2}^{\infty}\frac{n-1}{n}\int_{0}^{\ttot}\frac{\mexp{-\ftime(\tfir)}}{1-\mexp{-\ftime(\tfir)}}y^n\frac{n+1}{n-1}S_{n}(x)\mder \tfir\\
&=\int_0^{\ttot}\frac{\mexp{-\ftime(\tfir)}}{1-\mexp{-\ftime(\tfir)}}\sum_{n=2}^{\infty}
\left(1+\frac{1}{n}\right)
y^nS_{n}(x)\mder\tfir
\end{align*}

Let us compute
\[
\sum_{n=2}^{\infty}
\left(1+\frac{1}{n}\right)
y^nS_{n}(x)
=\sum_{n=2}^{\infty}
\left(1+\frac{1}{n}\right)
y^n\left(x(1+x)^{n-1}-2\int_0^1 xu(1+xu)^{n-1}\mder u\right).
\]
We shall use the identity, valid for $|a|<1$,
\[
\sum_{n=2}^{\infty}
\left(1+\frac{1}{n}\right)a^n
=
\frac{a^2}{1-a}
-a
-\log(1-a).
\]
Since
\[
0\leq y(1+xu)\leq y(1+x)=1-\mexp{-\ftime(\ttot)}<1,
\]
this identity can be applied with $a=y(1+x)$ and $a=y(1+xu)$, and the latter series can be integrated termwise. Hence
\begin{multline*}
\sum_{n=2}^{\infty}
\left(1+\frac{1}{n}\right)y^n S_{n}(x)
=
\frac{x}{1+x}
\left[
\frac{\left(y(1+x)\right)^2}{1-y(1+x)}
-y(1+x)
-\log\left(1-y(1+x)\right)
\right]
\\
-
2\int_0^1
\frac{xu}{1+xu}
\left[
\frac{\left(y(1+xu)\right)^2}{1-y(1+xu)}
-y(1+xu)
-\log\left(1-y(1+xu)\right)
\right]\mder u.
\end{multline*}
It remains to evaluate the integral. With the change of variable $z=y(1+xu)$,
we get
\begin{multline*}
\int_0^1
\frac{xu}{1+xu}
\left[
\frac{\left(y(1+xu)\right)^2}{1-y(1+xu)}
-y(1+xu)
-\log\left(1-y(1+xu)\right)
\right]\mder u
\\
=
\frac{1}{xy}
\int_y^{y(1+x)}
\frac{z-y}{z}
\left[
\frac{z^2}{1-z}
-z
-\log(1-z)
\right]\mder z .
\end{multline*}
A primitive of the integrand is
\[
-z^2+2yz+(y-z)\log(1-z)-y\operatorname{Li}_2(z).
\]
Therefore
\begin{multline*}
\int_0^1
\frac{xu}{1+xu}
\left[
\frac{\left(y(1+xu)\right)^2}{1-y(1+xu)}
-y(1+xu)
-\log\left(1-y(1+xu)\right)
\right]\mder u
\\
=
-xy
-\log\left(1-y(1+x)\right)
+
\frac{1}{x}
\left[
\operatorname{Li}_2(y)
-
\operatorname{Li}_2\left(y(1+x)\right)
\right].
\end{multline*}
Consequently,
\begin{multline*}
\sum_{n=2}^{\infty}
\left(1+\frac{1}{n}\right)y^n S_{n}(x)
=
\frac{x}{1+x}
\left[
\frac{\left(y(1+x)\right)^2}{1-y(1+x)}
-y(1+x)
-\log\left(1-y(1+x)\right)
\right]
\\
+
2xy
+
2\log\left(1-y(1+x)\right)
-
\frac{2}{x}
\left[
\operatorname{Li}_2(y)
-
\operatorname{Li}_2\left(y(1+x)\right)
\right].
\end{multline*}

Substituting back $x$ and $y$, we obtain
\begin{multline*}
\variance(\meanTrait[\treeRV(\ttot)]\mid \effRec[\treeRV(\ttot)](\ttot)\geq 1)
\\=
\sigma^2
\Bigg[
\ttot
-
\int_0^{\ttot}
\Bigg(
1+\frac{\ftime(\ttot)}{\mexp{\ftime(\ttot)}-1}
-
2\ftime(\ttot)
\frac{\mexp{-\ftime(\tfir)}}{1-\mexp{-\ftime(\tfir)}}
-
2
\frac{\mexp{\ftime(\ttot)}-\mexp{\ftime(\tfir)}}{(\mexp{\ftime(\tfir)}-1)^2}
\left[
\operatorname{Li}_2\left(1-\mexp{-(\ftime(\ttot)-\ftime(\tfir))}\right)
-
\operatorname{Li}_2\left(1-\mexp{-\ftime(\ttot)}\right)
\right]
\Bigg)
\mder\tfir
\Bigg]
\\=\sigma^2\left(-\frac{\ttot\ftime(\ttot)}{\mexp{\ftime(\ttot)}-1}+2\int_0^{\ttot}\left(
\frac{\ftime(\ttot)}{\mexp{\ftime(\tfir)}-1}+\frac{\mexp{\ftime(\ttot)}-\mexp{\ftime(\tfir)}}{(\mexp{\ftime(\tfir)}-1)^2}
\left[
\operatorname{Li}_2\left(1-\mexp{-(\ftime(\ttot)-\ftime(\tfir))}\right)
-
\operatorname{Li}_2\left(1-\mexp{-\ftime(\ttot)}\right)
\right]\right)\mder\tfir\right).
\end{multline*}

Let us set
\[
H_{\ttot}(\tfir)
=
\frac{\ftime(\ttot)}{\mexp{\ftime(\tfir)}-1}
+
\frac{\mexp{\ftime(\ttot)}-\mexp{\ftime(\tfir)}}{
\left(\mexp{\ftime(\tfir)}-1\right)^2}
\left[
\operatorname{Li}_2\left(1-\mexp{-(\ftime(\ttot)-\ftime(\tfir))}\right)-
\operatorname{Li}_2\left(1-\mexp{-\ftime(\ttot)}\right)
\right].
\]

Let us show that $H_{\ttot}$ can be continuously extended at
$0$. This implies in particular that it is integrable on
$[0,\ttot]$.

Since
\[
\frac{d}{du}
\operatorname{Li}_2\left(1-\mexp{-u}\right)
=
\frac{u}{\mexp{u}-1} \mbox{ and } \frac{\mder^{2}}{\mder u^{2}}
\operatorname{Li}_2\left(1-\mexp{-u}\right)
=
\frac{\mexp{u}(1-u)-1}{(\mexp{u}-1)^{2}},
\]

the Taylor expansion of $\operatorname{Li}_2\left(1-\mexp{-u}\right)$ at $\ftime(\ttot)$ gives
\[
\operatorname{Li}_2\left(1-\mexp{-(\ftime(\ttot)-\ftime(\tfir))}\right)-\operatorname{Li}_2\left(1-\mexp{-\ftime(\ttot)}\right)
=
-\frac{\ftime(\ttot)}{\mexp{\ftime(\ttot)}-1}\ftime(\tfir)
-
\frac{
\mexp{\ftime(\ttot)}(\ftime(\ttot)-1)+1
}{
2\left(\mexp{\ftime(\ttot)}-1\right)^2
}
\ftime(\tfir)^2
+
o\left(\ftime(\tfir)^2\right).
\]
Moreover,
\[
\mexp{u}-1 = u\left(1+\frac{u}{2}+o(u)\right) \mbox{, thus } \frac{1}{(\mexp{u}-1)^{2}} = \frac{1}{u^{2}}(1-u+o(u)).
\]
It follows that 
\[
\frac{\mexp{\ftime(\ttot)}-\mexp{\ftime(\tfir)}}{
\left(\mexp{\ftime(\tfir)}-1\right)^2}
=
\frac{\mexp{\ftime(\ttot)}-1}{\ftime(\tfir)^2}
-
\frac{\mexp{\ftime(\ttot)}}{\ftime(\tfir)}
+
o\left(\frac{1}{\ftime(\tfir)}\right),
\]
and
\[
\frac{\ftime(\ttot)}{\mexp{\ftime(\tfir)}-1}
=
\frac{\ftime(\ttot)}{\ftime(\tfir)}
-
\frac{\ftime(\ttot)}{2}
+
o(1).
\]
Substituting these expansions in the definition of $H_{\ttot}(\tfir)$, the
terms in $1/\ftime(\tfir)$ cancel, 
\begin{align*}
H_{\ttot}(\tfir)
&=
\frac{\ftime(\ttot)}{\ftime(\tfir)}
-
\frac{\ftime(\ttot)}{2}
-
\left(\frac{\mexp{\ftime(\ttot)}-1}{\ftime(\tfir)^2}
-
\frac{\mexp{\ftime(\ttot)}}{\ftime(\tfir)}
\right)
\left[\frac{\ftime(\ttot)}{\mexp{\ftime(\ttot)}-1}\ftime(\tfir)
+
\frac{
\mexp{\ftime(\ttot)}(\ftime(\ttot)-1)+1
}{
2\left(\mexp{\ftime(\ttot)}-1\right)^2
}
\ftime(\tfir)^2
\right] + o(1)\\
&=\frac{1}{2}+
\frac{\ftime(\ttot)}{2(\mexp{\ftime(\ttot)}-1)}+o(1)
.
\end{align*}

This proves that the integrand is well defined as an improper integral at
$\tfir=0$, and completes the proof.

\section{Proof of Propositions~\ref{prop:closednon} and~\ref{prop:closedcrit}}
\subsection{Preliminaries}
Proposition~\ref{prop:expect} yields
\begin{equation}\label{eq:reexp}
\esp\left(\mavar{\treeRV(\ttot)}\right)
=
\sigma^{2}\pSurvival(0,\ttot)(1-\mexp{-\ftime(\ttot)})\left(\ttot -\frac{2}{\mexp{\ftime(\ttot)}-1}I(\ttot)\right),
\end{equation}
where
\[
I(\ttot) = \int_0^{\ttot}\frac{(\mexp{\ftime(\ttot)}-\mexp{\ftime(\tfir)})\left(1-\mexp{\ftime(\tfir)}+\ftime(\tfir)\mexp{\ftime(\tfir)}\right)}{\left(\mexp{\ftime(\tfir)}-1\right)^{2}}\mder\tfir.
\]
Since 
$\frac{\mder}{\mder u}\left(\frac{u}{\mexp{u}-1}\right)=-\frac{1-\mexp{u}+u\mexp{u}}{(\mexp{u}-1)^2}$, we have that 
\[
I(\ttot) = -\int_0^{\ttot}\frac{\mexp{\ftime(\ttot)}-\mexp{\ftime(\tfir)}}{\ftime'(\tfir)}
\frac{d}{d\tfir}\left(\frac{\ftime(\tfir)}{\mexp{\ftime(\tfir)}-1}\right)\mder\tfir.
\]
Integrating by parts leads to 
\[
I(\ttot) = -\left[\frac{\mexp{\ftime(\ttot)}-\mexp{\ftime(\tfir)}}{\ftime'(\tfir)}\times\frac{\ftime(\tfir)}{\mexp{\ftime(\tfir)}-1}\right]_{0}^{\ttot}+\int_0^{\ttot}\frac{d}{d\tfir}\left(\frac{\mexp{\ftime(\ttot)}-\mexp{\ftime(\tfir)}}{\ftime'(\tfir)}\right)
\frac{\ftime(\tfir)}{\mexp{\ftime(\tfir)}-1}\mder\tfir.
\]
Since $\ftime(0) = 0$ and $\lim_{a\rightarrow 0}\frac{a}{\mexp{a}-1} = 1$, we get
\begin{equation}\label{eq:integral}
I(\ttot) = \frac{\mexp{\ftime(\ttot)}-1}{\ftime'(0)}+\int_0^{\ttot}\frac{d}{d\tfir}\left(\frac{\mexp{\ftime(\ttot)}-\mexp{\ftime(\tfir)}}{\ftime'(\tfir)}\right)
\frac{\ftime(\tfir)}{\mexp{\ftime(\tfir)}-1}\mder\tfir.
\end{equation}

\subsection{Non-critical case - Proposition~\ref{prop:closednon}}
We have 
\[
\ftime(\tfir)
=
(\birth-\death)\tfir-\log\left(
\frac{\birth-\death \mexp{-(\birth-\death)(\ttot-\tfir)}}{\birth-\death \mexp{-(\birth-\death)\ttot}}
\right),
\]
Therefore
\[
\mexp{\ftime(\tfir)}
=
\frac{\mexp{(\birth-\death)\tfir}\left(\birth-\death \mexp{-(\birth-\death)\ttot}\right)}
{\birth-\death \mexp{-(\birth-\death)(\ttot-\tfir)}},\quad \mexp{\ftime(\ttot)}
=
\frac{\birth \mexp{(\birth-\death)\ttot}-\death}{(\birth-\death)},
\qquad
\ftime'(\tfir)
=
\frac{\birth (\birth-\death)}{\birth-\death \mexp{-(\birth-\death)(\ttot-\tfir)}}, 
\qquad
\ftime'(0)
=
\frac{\birth (\birth-\death)}{\birth-\death \mexp{-(\birth-\death)\ttot}},
\]
and
\[\frac{\mexp{\ftime(\ttot)}-\mexp{\ftime(\tfir)}}{\ftime'(\tfir)}=\frac{\left(\birth-\death \mexp{-(\birth-\death)\ttot}\right)\left(\mexp{(\birth-\death)\ttot}-\mexp{(\birth-\death)\tfir}\right)}{(\birth-\death)^2},\quad
\frac{d}{d\tfir}\left(
\frac{\mexp{\ftime(\ttot)}-\mexp{\ftime(\tfir)}}{\ftime'(\tfir)}
\right)
=
-\frac{\left(\birth-\death \mexp{-(\birth-\death)\ttot}\right)\mexp{(\birth-\death)\tfir}}{(\birth-\death)}.
\]

Substituting in Equation~\ref{eq:integral} gives
\[
I(\ttot)
=
\frac{\left(\birth-\death \mexp{-(\birth-\death)\ttot}\right)\left(\mexp{(\birth-\death)\ttot}-1\right)}{(\birth-\death)^2}
-
\frac{\birth-\death \mexp{-(\birth-\death)\ttot}}{\birth (\birth-\death)}
\int_0^{\ttot}
\frac{\mexp{(\birth-\death)\tfir}\left(\birth-\death \mexp{-(\birth-\death)(\ttot-\tfir)}\right)}
{\mexp{(\birth-\death)\tfir}-1}
\log\left(
\frac{\mexp{(\birth-\death)\tfir}\left(\birth-\death \mexp{-(\birth-\death)\ttot}\right)}
{\birth-\death \mexp{-(\birth-\death)(\ttot-\tfir)}}
\right)\mder \tfir.
\]

The change of variable $x=\mexp{(\birth-\death)\tfir}, d\tfir=\frac{dx}{(\birth-\death)x}$ leads to
\[
I(\ttot)
=
\frac{\left(\birth-\death \mexp{-(\birth-\death)\ttot}\right)\left(\mexp{(\birth-\death)\ttot}-1\right)}{(\birth-\death)^2}
-
\frac{\birth-\death \mexp{-(\birth-\death)\ttot}}{\birth (\birth-\death)^2}
\int_1^{\mexp{(\birth-\death)\ttot}}
\frac{\birth-\death \mexp{-(\birth-\death)\ttot}x}{x-1}
\log\left(
\frac{\left(\birth-\death \mexp{-(\birth-\death)\ttot}\right)x}
{\birth-\death \mexp{-(\birth-\death)\ttot}x}
\right)\mder x.
\]

We have
\begin{multline*}
\int_1^{\mexp{(\birth-\death)\ttot}}
\frac{\birth-\death \mexp{-(\birth-\death)\ttot}x}{x-1}
\log\left(
\frac{\left(\birth-\death \mexp{-(\birth-\death)\ttot}\right)x}
{\birth-\death \mexp{-(\birth-\death)\ttot}x}
\right)\mder x = \\
-\death \mexp{-(\birth-\death)\ttot}\int_1^{\mexp{(\birth-\death)\ttot}}
\log\left(
\frac{\left(\birth-\death \mexp{-(\birth-\death)\ttot}\right)x}
{\birth-\death \mexp{-(\birth-\death)\ttot}x}
\right)\mder x + 
(\birth-\death \mexp{-(\birth-\death)\ttot})\int_1^{\mexp{(\birth-\death)\ttot}}
\frac{1}{x-1}
\log\left(
\frac{\left(\birth-\death \mexp{-(\birth-\death)\ttot}\right)x}
{\birth-\death \mexp{-(\birth-\death)\ttot}x}
\right)\mder x.
\end{multline*}
For the first term, we use
\begin{multline*}
\int_1^{\mexp{(\birth-\death)\ttot}}
\log\left(
\frac{\left(\birth-\death \mexp{-(\birth-\death)\ttot}\right)x}
{\birth-\death \mexp{-(\birth-\death)\ttot}x}
\right)\mder x
=\\
\int_1^{\mexp{(\birth-\death)\ttot}}\log\left(\birth-\death \mexp{-(\birth-\death)\ttot}\right)\mder x
+
\int_1^{\mexp{(\birth-\death)\ttot}}\log (x)\mder dx
-
\int_1^{\mexp{(\birth-\death)\ttot}}\log\left(\birth-\death \mexp{-(\birth-\death)\ttot}x\right)\mder x.
\end{multline*}
which gives
\[
-\death \mexp{-(\birth-\death)\ttot}
\int_1^{\mexp{(\birth-\death)\ttot}}
\log\left(
\frac{\left(\birth-\death \mexp{-(\birth-\death)\ttot}\right)x}
{\birth-\death \mexp{-(\birth-\death)\ttot}x}
\right)\mder x
=
(\birth-\death)\log\left(
\frac{\birth-\death \mexp{-(\birth-\death)\ttot}}{(\birth-\death)}
\right)
-
\death (\birth-\death)\ttot.
\]

For the second term, we write that
\begin{multline*}
\int_1^{\mexp{(\birth-\death)\ttot}}
\frac{\birth-\death \mexp{-(\birth-\death)\ttot}}{x-1}
\log\left(
\frac{\left(\birth-\death \mexp{-(\birth-\death)\ttot}\right)x}
{\birth-\death \mexp{-(\birth-\death)\ttot}x}
\right)\mder x
=\\
\left(\birth-\death \mexp{-(\birth-\death)\ttot}\right)
\left(\int_1^{\mexp{(\birth-\death)\ttot}}
\frac{\log (x)}{x-1}\mder x
-
\int_1^{\mexp{(\birth-\death)\ttot}}
\frac{
\log\left(
1-\frac{\death \mexp{-(\birth-\death)\ttot}}{\birth-\death \mexp{-(\birth-\death)\ttot}}(x-1)
\right)
}{x-1}\mder x\right).
\end{multline*}

Using Equation~\ref{eq:dilog:int}, we obtain
\[
\operatorname {Li} _{2}(y) = -\int _{0}^{y}{\frac {\ln (1-u)}{u}}\mder u\mbox{ or, equivalently }\operatorname {Li} _{2}(1-z) = \int _{1}^{z}{\frac {\ln u}{1-u}}\mder u = -\int _{1}^{z}{\frac {\ln u}{u-1}}\mder u.
\]

With the change of variable
$u=\frac{\death \mexp{-(\birth-\death)\ttot}}{\birth-\death \mexp{-(\birth-\death)\ttot}}(x-1)$ in the second integral,
we get
\[
\int_1^{\mexp{(\birth-\death)\ttot}}
\frac{\birth-\death \mexp{-(\birth-\death)\ttot}}{x-1}
\log\left(
\frac{\left(\birth-\death \mexp{-(\birth-\death)\ttot}\right)x}
{\birth-\death \mexp{-(\birth-\death)\ttot}x}
\right)\mder x
=
\left(\birth-\death \mexp{-(\birth-\death)\ttot}\right)
\left[
-\operatorname{Li}_2\left(1-\mexp{(\birth-\death)\ttot}\right)
+
\operatorname{Li}_2\left(
\frac{\death\left(1-\mexp{-(\birth-\death)\ttot}\right)}
{\birth-\death \mexp{-(\birth-\death)\ttot}}
\right)
\right].
\]

Combining formulas above, we finally obtain
\[
I(\ttot)
=
\frac{\left(\birth-\death \mexp{-(\birth-\death)\ttot}\right)
\left(\mexp{(\birth-\death)\ttot}-1\right)}
{(\birth-\death)^2}
-
\frac{\birth-\death \mexp{-(\birth-\death)\ttot}}
{\birth(\birth-\death)}
\log\left(
\frac{\birth-\death \mexp{-(\birth-\death)\ttot}}
{\birth-\death}
\right)
\]
\[
\qquad
+
\frac{\death\ttot\left(\birth-\death \mexp{-(\birth-\death)\ttot}\right)}
{\birth(\birth-\death)}
+
\frac{\left(\birth-\death \mexp{-(\birth-\death)\ttot}\right)^2}
{\birth(\birth-\death)^2}
\left[
\operatorname{Li}_2\left(1-\mexp{(\birth-\death)\ttot}\right)
-
\operatorname{Li}_2\left(
\frac{\death\left(1-\mexp{-(\birth-\death)\ttot}\right)}
{\birth-\death \mexp{-(\birth-\death)\ttot}}
\right)
\right].
\]
Proposition~\ref{prop:closednon} then follows from Equation~\ref{eq:reexp}.

\subsection{Critical case - Proposition~\ref{prop:closedcrit}}
In the critical case, we have
\[
\ftime(\tfir)
=
\log\left(
\frac{1+\birth\ttot}{1+\birth(\ttot-\tfir)}
\right).
\]
Hence
\[
\mexp{\ftime(\tfir)}
=
\frac{1+\birth\ttot}{1+\birth(\ttot-\tfir)}, 
\qquad
\mexp{\ftime(\ttot)}=1+\birth\ttot,
\qquad
\ftime'(\tfir)
=
\frac{\birth}{1+\birth(\ttot-\tfir)},
\qquad
\ftime'(0)=\frac{\birth}{1+\birth\ttot}.
\]
Therefore
\[
\frac{\mexp{\ftime(\ttot)}-\mexp{\ftime(\tfir)}}{\ftime'(\tfir)}
=
\frac{
(1+\birth\ttot)
-
\frac{1+\birth\ttot}{1+\birth(\ttot-\tfir)}
}{
\frac{\birth}{1+\birth(\ttot-\tfir)}
}
=
(1+\birth\ttot)(\ttot-\tfir),
\quad
\frac{d}{d\tfir}
\left(
\frac{\mexp{\ftime(\ttot)}-\mexp{\ftime(\tfir)}}{\ftime'(\tfir)}
\right)
=
-(1+\birth\ttot).
\]

Equation~\ref{eq:integral} becomes
\[
I(\ttot)
=
\ttot(1+\birth\ttot)
-
(1+\birth\ttot)
\int_0^{\ttot}
\frac{1+\birth(\ttot-\tfir)}{\birth \tfir}
\log\left(
\frac{1+\birth\ttot}{1+\birth(\ttot-\tfir)}
\right)\mder \tfir.
\]

The change of variable $x=\frac{1+\birth(\ttot-\tfir)}{1+\birth\ttot}$, $\mder \tfir=-\frac{1+\birth\ttot}{\birth}\,\mder x$ leads to
\[
I(\ttot)
=
\ttot(1+\birth\ttot)
-
\frac{(1+\birth\ttot)^2}{\birth}
\int_{1/(1+\birth\ttot)}^1
\frac{x}{1-x}\log\left(\frac1x\right)\mder x.
\]

We have 
\begin{align*}
\int_{1/(1+\birth\ttot)}^1\frac{x}{1-x}\log\left(\frac1x\right)\mder x
&=
\int_{1/(1+\birth\ttot)}^1\left(\log(x) +\frac{\log (x)}{1-x}\right)\mder x\\
&=\left[x\log(x)-x-\operatorname{Li}_2(1-x)\right]_{1/(1+\birth\ttot)}^1\\
&=-1+\frac{\log(1+\birth\ttot)+1}{1+\birth\ttot}
+\operatorname{Li}_2\left(\frac{\birth\ttot}{1+\birth\ttot}\right).
\end{align*}

Finally, we obtain
\[
I(\ttot)
=
2\ttot(1+\birth\ttot)
-
\frac{1+\birth\ttot}{\birth}\log(1+\birth\ttot)
-
\frac{(1+\birth\ttot)^2}{\birth}
\operatorname{Li}_2\left(\frac{\birth\ttot}{1+\birth\ttot}\right).
\]
Proposition~\ref{prop:closedcrit} then follows from Equation~\ref{eq:reexp}.

\section{Proof of Proposition~\ref{prop:asymp}}
\subsection{Supercritical case}
The closed formula can be written as
\begin{multline*}\esp(\mavar{\treeRV(\ttot)})
=
\frac{\sigma^2(\birth-\death)}{\birth-\death\mexp{-(\birth-\death)\ttot}}
\Bigg[
\ttot-\ttot \mexp{-(\birth-\death)\ttot}\left(
\frac{\birth-\death}{\birth-\death \mexp{-(\birth-\death)\ttot}}
+
\frac{2\death}{\birth}
\right)
-
\frac{2\left(1-\mexp{-(\birth-\death)\ttot}\right)}{\birth-\death}
\\+
\frac{2\mexp{-(\birth-\death)\ttot}}{\birth}
\log\left(
\frac{\birth-\death \mexp{-(\birth-\death)\ttot}}{\birth-\death}
\right)
-
\frac{2\mexp{-(\birth-\death)\ttot}\left(\birth-\death \mexp{-(\birth-\death)\ttot}\right)}{\birth(\birth-\death)}
\left(
\operatorname{Li}_2\left(1-\mexp{(\birth-\death)\ttot}\right)
-
\operatorname{Li}_2\left(
\frac{\death\left(1-\mexp{-(\birth-\death)\ttot}\right)}
{\birth-\death \mexp{-(\birth-\death)\ttot}}
\right)
\right)
\Bigg].
\end{multline*}
From Equation~\ref{eq:dilogrefA}, we get
\[
\operatorname{Li}_2\left(1-\mexp{(\birth-\death)\ttot}\right) = -\frac{\pi^{2}}{6}-\frac{\log^{2}(\mexp{(\birth-\death)\ttot}-1)}{2}-\operatorname{Li}_2\left(\frac{1}{1-\mexp{(\birth-\death)\ttot}}\right).
\]
Therefore, if $\birth > \death$, the last
term inside the brackets of the closed formula of $\esp(\mavar{\treeRV(\ttot)})$ displayed above tends to $0$ as $\ttot\to\infty$. All terms in the expression inside the brackets have finite limits as $\ttot\to\infty$, except for the
leading $\ttot$ term. Moreover, we have
\[
\frac{\sigma^2(\birth-\death)}{\birth-\death\mexp{-(\birth-\death)\ttot}} = \frac{\sigma^2(\birth-\death)}{\birth}\left(1+\frac{\death}{\birth}\mexp{-(\birth-\death)\ttot}\right)+o(\mexp{-(\birth-\death)\ttot}).
\]
Altogether, this yields 
\[
\esp(\mavar{\treeRV(\ttot)})=\frac{\sigma^2(\birth-\death)}{\birth}\left(\ttot-\frac{2}{\birth-\death}\right)+o(1).
\]
\subsection{Critical case}
Using elementary limits term by term, together with the continuity of $\operatorname{Li}_2$ at $1$ and the fact that $\operatorname{Li}_2(1) = \frac{\pi^{2}}{6}$ (Eq.~\ref{eq:dilog:vun}), we get
\begin{align*}
\lim_{\ttot\to\infty}\esp(\mavar{\treeRV(\ttot)}) 
&= \lim_{\ttot\to\infty} \sigma^2
\left[
\frac{\birth \ttot^2}{(1+\birth \ttot)^{2}}
-\frac{4\ttot}{1+\birth \ttot}
+\frac{2}{\birth(1+\birth \ttot)}\log(1+\birth \ttot)
+\frac{2}{\birth}
\operatorname{Li}_2\left(\frac{\birth \ttot}{1+\birth \ttot}\right)
\right]\\
&= \frac{\sigma^{2}}{\birth}\left(\frac{\pi^{2}}{3}-3\right)
\end{align*}

\subsection{Subcritical case}
The closed formula of Proposition~\ref{prop:closednon} can be written as
\begin{multline*}\esp(\mavar{\treeRV(\ttot)})
=
\frac{\sigma^2(\death-\birth)\mexp{-(\death-\birth)\ttot}}{\death-\birth\mexp{-(\death-\birth)\ttot}}
\Bigg[
\ttot\left(
-\frac{\birth\left(\mexp{-(\death-\birth)\ttot}-1\right)}{\death-\birth\mexp{-(\death-\birth)\ttot}}
-
\frac{2\death}{\birth\mexp{-(\death-\birth)\ttot}}
\right)
-
\frac{2\left(1-\mexp{-(\death-\birth)\ttot}\right)}{(\death-\birth)\mexp{-(\death-\birth)\ttot}}
\\+
\frac{2}{\birth\mexp{-(\death-\birth)\ttot}}
\log\left(
\frac{\death-\birth\mexp{-(\death-\birth)\ttot}}{(\death-\birth)\mexp{-(\death-\birth)\ttot}}
\right)
-
\frac{2\left(\death -\birth\mexp{-(\death-\birth)\ttot}\right)}{\birth(\death-\birth)\mexp{-2(\death-\birth)\ttot}}
\left(
\operatorname{Li}_2\left(1-\mexp{-(\death-\birth)\ttot}\right)
-
\operatorname{Li}_2\left(
\frac{\death\left(1-\mexp{-(\death-\birth)\ttot}\right)}
{\death-\birth\mexp{-(\death-\birth)\ttot}}
\right)
\right)
\Bigg].
\end{multline*}
Setting $x = \mexp{-(\death-\birth)\ttot}$, this becomes
\begin{multline*}\esp(\mavar{\treeRV(\ttot)})
=
\frac{\sigma^2(\death-\birth) x}{\death-\birth x}
\Bigg[
\frac{\log(x)}{\death-\birth}\left(
\frac{\birth\left( x-1\right)}{\death-\birth x}
+
\frac{2\death}{\birth x}
\right)
-
\frac{2\left(1-x\right)}{(\death-\birth) x}
\\+
\frac{2}{\birth x}
\log\left(
\frac{\death-\birth x}{(\death-\birth) x}
\right)
-
\frac{2\left(\death -\birth x\right)}{\birth(\death-\birth)x^{2}}
\left(
\operatorname{Li}_2\left(1- x\right)
-
\operatorname{Li}_2\left(
\frac{\death\left(1- x\right)}
{\death-\birth x}
\right)
\right)
\Bigg].
\end{multline*}
We first derive an expansion of 
\[
\operatorname{Li}_2\left(1-x\right)
-
\operatorname{Li}_2\left(
\frac{\death\left(1- x\right)}
{\death-\birth x}
\right)
=\operatorname{Li}_2\left(1-x\right)
-
\operatorname{Li}_2\left(1-f(x)
\right),
\]
with $f(x) = \frac{(\death-\birth)x}
{\death-\birth x}$.

We have 
\begin{equation*}
f(x) = \frac{(\death-\birth) x}{\death}\times\frac{1}{1-\frac{\birth x}{\death}}
=\frac{(\death-\birth) x}{\death}\left(1+\frac{\birth x}{\death}+\frac{\birth^{2} x^{2}}{\death^{2}}+o(x^{2})\right).
\end{equation*}
Therefore
\begin{equation*}
\log(f(x)) = \log\left(\frac{\death-\birth}{\death}\right)+\log(x)-\log\left(1-\frac{\birth x}{\death}\right)
=\log\left(\frac{\death-\birth}{\death}\right)+\log(x)+\frac{\birth x}{\death}+o(x).
\end{equation*}

Equation~\ref{eq:dilog:dev} gives
\begin{align*}
\operatorname{Li}_2(1-x) 
 &= \frac{\pi^{2}}{6}+x(\log(x)-1)+\frac{x^{2}}{2}\left(\log(x)-\frac{1}{2}\right)+o(x^{2})
 \end{align*}
It follows that
\begin{align*}
\operatorname{Li}_2(1-f(x)) 
 &= \frac{\pi^{2}}{6}+f(x)(\log(f(x))-1)+\frac{f(x)^{2}}{2}\left(\log(f(x))-\frac{1}{2}\right)+o(f(x)^{2})\\
&=\frac{\pi^{2}}{6}
+x\frac{(\death-\birth)}{\death}\left(\log\left(\frac{\death-\birth}{\death}\right)-1\right)
+x\log(x)\frac{(\death-\birth)}{\death}
\\&+x^{2}\left(\frac{(\death-\birth)}{\death}\log\left(\frac{\death-\birth}{\death}\right)\frac{\birth}{\death}+\frac{1}{2}\left(\frac{(\death-\birth)}{\death}\right)^{2}\left(\log\left(\frac{\death-\birth}{\death}\right)-\frac{1}{2}\right)\right)
\\&+x^{2}\log(x)\left(\frac{(\death-\birth)}{\death}\frac{\birth}{\death}+\frac{1}{2}\left(\frac{(\death-\birth)}{\death}\right)^{2}\right)+o(x^{2}).
\end{align*}

We obtain
\begin{multline*}
\operatorname{Li}_2\left(1-x\right)
-
\operatorname{Li}_2\left(1-
f(x)\right) 
=
-x\left(\frac{\birth}{\death}+\frac{(\death-\birth)}{\death}\log\left(\frac{\death-\birth}{\death}\right)\right)
+x\log(x)\frac{\birth}{\death}
\\-x^{2}\left(\frac{\death^2-\birth^2}{2\death^2}
\log\left(\frac{\death-\birth}{\death}\right)
+
\frac{\birth(2\death-\birth)}{4\death^2}\right)
+x^{2}\log(x)\frac{1}{2}\frac{\birth^{2}}{\death^{2}}+o(x^{2}).
\end{multline*}
Returning to the expression of $\esp(\mavar{\treeRV(\ttot)})$, after substitution of the expansion above, the terms in $\log(x)/x$, $1/x$, and $\log(x)$ cancel, while the remaining $x$-dependent terms are absorbed into the $o(x)$ term. We finally obtain
\begin{equation*}\esp(\mavar{\treeRV(\ttot)})
=\frac{\sigma^2}{\death^{2}}
\Bigg[
\frac{(\death-\birth)^{2}}{\birth}\log\left(\frac{\death-\birth}{\death}\right)+\death-\frac{\birth}{2}\Bigg]x+o(x).
\end{equation*}
Substituting back $x$ yields
\[
\esp(\mavar{\treeRV(\ttot)}) = \frac{\sigma^2}{\death^{2}}\left(\frac{(\death-\birth)^{2}}{\birth}\log\left(\frac{\death-\birth}{\death}\right)+\death-\frac{\birth}{2}\right)\mexp{-(\death-\birth)\ttot}+o(\mexp{-(\death-\birth)\ttot}).
\]
This proves the result.

\section{Proof of Corollary~\ref{cor:asympcond}}
Since, by convention, the empirical variance is zero when no lineage is extant
at time $\ttot$, we have
\[
\esp\left(
\mavar{\treeRV(\ttot)}
\mid \effRec[\treeRV(\ttot)](\ttot)\geq 1
\right)
=
\frac{
\esp\left(\mavar{\treeRV(\ttot)}\right)
}{
\prob\left(\effRec[\treeRV(\ttot)](\ttot)\geq 1\right)
}.
\]

If $\birth\neq\death$, we have
\[
\frac{1}{\prob\left(\effRec[\treeRV(\ttot)](\ttot)\geq 1\right)}
=
\frac{\birth}{\birth-\death}-\frac{\death}{\birth-\death}\mexp{-(\birth-\death)\ttot}.
\]
If $\birth>\death$, using Proposition~\ref{prop:asymp}, we obtain
\[
\esp\left(
\mavar{\treeRV(\ttot)}
\mid \effRec[\treeRV(\ttot)](\ttot)\geq 1
\right)
=
\sigma^2\ttot-\frac{2\sigma^2}{\birth-\death}+o(1).
\]

If $\birth=\death$, the survival probability is
\[
\prob\left(\effRec[\treeRV(\ttot)](\ttot)\geq 1\right)
=
\frac{1}{1+\birth\ttot}.
\]
Together with Proposition~\ref{prop:asymp}, this gives
\[
\esp\left(
\mavar{\treeRV(\ttot)}
\mid \effRec[\treeRV(\ttot)](\ttot)\geq 1
\right)
=
\sigma^{2}\ttot\left(\frac{\pi^{2}}{3}-3\right)+o(\ttot).
\]
In order to get a more precise result, we go back to
\[
\esp\left(
\mavar{\treeRV(\ttot)}
\mid \effRec[\treeRV(\ttot)](\ttot)\geq 1
\right) =\sigma^2
\left[
\frac{\birth \ttot^2}{1+\birth \ttot}
-4\ttot
+\frac{2}{\birth}\log(1+\birth \ttot)
+\frac{2(1+\birth \ttot)}{\birth}
\operatorname{Li}_2\left(\frac{\birth \ttot}{1+\birth \ttot}\right)
\right].
\]
Setting $x=\frac{1}{1+\birth \ttot}$, we get
\[
\frac{2}{\birth}\log(1+\birth \ttot)
+\frac{2(1+\birth \ttot)}{\birth}
\operatorname{Li}_2\left(\frac{\birth \ttot}{1+\birth \ttot}\right) = \frac{2}{\birth}\left(-\log(x)+\frac{\operatorname{Li}_2(1-x)}{x}\right).
\]

From Equation~\ref{eq:dilog:dev}, we obtain
\[
\frac{2}{\birth}\log(1+\birth \ttot)
+\frac{2(1+\birth \ttot)}{\birth}
\operatorname{Li}_2\left(\frac{\birth \ttot}{1+\birth \ttot}\right) = \frac{2}{\birth}\left(\frac{\pi^{2}}{6x}- 1\right)+o(1).
\]
Substituting back $x$, using
\[
\frac{\birth \ttot^2}{1+\birth \ttot}
=
\ttot-\frac{1}{\birth}+o(1),
\]
and putting everything together, we obtain
 \[
\esp\left(
\mavar{\treeRV(\ttot)}
\mid \effRec[\treeRV(\ttot)](\ttot)\geq 1
\right)
=
\sigma^2\left(\frac{\pi^2}{3}-3\right)\left(\ttot+\frac{1}{\birth}\right)
+
o(1).
\]
Finally, if $\birth<\death$, multiplying the subcritical equivalent of Proposition~\ref{prop:asymp} by the inverse of the survival probability gives
\[
\esp\left(
\mavar{\treeRV(\ttot)}
\mid \effRec[\treeRV(\ttot)](\ttot)\geq 1
\right)
=
\sigma^{2}\frac{\death}{\death-\birth}\Upsilon_{\birth,\death}+o(1).
\]

This proves the result.
\end{appendix}

\bibliographystyle{plainnat}

\bibliography{PhyloTraitDynamics.bib}

@article{Bartoszek2014,
title = {Quantifying the effects of anagenetic and cladogenetic evolution},
journal = {Mathematical Biosciences},
volume = {254},
pages = {42-57},
year = {2014},
issn = {0025-5564},
doi = {https://doi.org/10.1016/j.mbs.2014.06.002},
url = {https://www.sciencedirect.com/science/article/pii/S0025556414001084},
author = {Krzysztof Bartoszek}
}

@article{Bartoszek2015,
title = {A consistent estimator of the evolutionary rate},
journal = {Journal of Theoretical Biology},
volume = {371},
pages = {69-78},
year = {2015},
issn = {0022-5193},
doi = {https://doi.org/10.1016/j.jtbi.2015.01.019},
url = {https://www.sciencedirect.com/science/article/pii/S0022519315000284},
author = {Krzysztof Bartoszek and Serik Sagitov}
}

@inproceedings{Bastide2024,
  title        = {Evolutionary Models of Continuous Traits},
  author       = {Bastide, Paul and Mariadassou, Mahendra and Robin, Stéphane},
  year         = {2024},
  month        = {May},
  booktitle    = {Models and Methods for Biological Evolution},
  publisher    = {ISTE Wiley},
  address      = {London},
  pages        = {39--78},
  editor       = {Didier, G. and Guindon, S.}
}

@book{Casella2002,
  author = {Casella, George and Berger, Roger L.},
  publisher = {Duxbury Press},
  edition = {2nd},
  address = {Pacific Grove},
  title = {Statistical inference},
  year = 2002
}

@article{Crawford2013,
    author = {Crawford, Forrest W. and Suchard, Marc A.},
    title = {Diversity, Disparity, and Evolutionary Rate Estimation for Unresolved {Y}ule Trees},
    journal = {Systematic Biology},
    volume = {62},
    number = {3},
    pages = {439-455},
    year = {2013},
    month = {03},
    issn = {1063-5157},
    doi = {10.1093/sysbio/syt010},
    url = {https://doi.org/10.1093/sysbio/syt010}
}

@article{Davies1980,
    author = {Davies, Robert B.},
    title = {The Distribution of a Linear Combination of $\chi^{2}$ Random Variables},
    journal = {Journal of the Royal Statistical Society Series C: Applied Statistics},
    volume = {29},
    number = {3},
    pages = {323-333},
    year = {1980},
    month = {12},
    issn = {0035-9254},
    doi = {10.2307/2346911},
    url = {https://doi.org/10.2307/2346911}
 }

@article{Duchesne2010,
    title = {Computing the Distribution of Quadratic Forms: Further
      Comparisons Between the {L}iu-{T}ang-{Z}hang Approximation and Exact
      Methods},
    author = {Pierre Duchesne and Pierre {Lafaye De Micheaux}},
    journal = {Computational Statistics and Data Analysis},
    year = {2010},
    pages = {858-862},
    volume = {54},
}

@article{Felsenstein1985,
  author  = {Felsenstein, Joseph},
  title   = {Phylogenies and the Comparative Method},
  journal = {The American Naturalist},
  year    = {1985},
  volume  = {125},
  number  = {1},
  pages   = {1--15},
  doi     = {10.1086/284325}
}

@article{Foote1993, title={Discordance and concordance between morphological and taxonomic diversity}, volume={19}, DOI={10.1017/S0094837300015864}, number={2}, journal={Paleobiology}, author={Foote, Mike}, year={1993}, pages={185–204}}

@article{Foote1997,
   author = "Foote, Mike",
   title = "The Evolution of Morphological Diversity", 
   journal= "Annual Review of Ecology, Evolution, and Systematics",
   year = "1997",
   volume = "28",
   pages = "129-152",
   doi = "https://doi.org/10.1146/annurev.ecolsys.28.1.129",
   url = "https://www.annualreviews.org/content/journals/10.1146/annurev.ecolsys.28.1.129",
   publisher = "Annual Reviews",
   issn = "1545-2069",
   type = "Journal Article"
  }

@article{Grafen1989,
  author  = {Grafen, Alan},
  title   = {The Phylogenetic Regression},
  journal = {Philosophical Transactions of the Royal Society of London. Series B, Biological Sciences},
  year    = {1989},
  volume  = {326},
  number  = {1233},
  pages   = {119--157},
  doi     = {10.1098/rstb.1989.0106}
}

@article{Harmon2003,
  author  = {Harmon, Luke J. and Schulte, James A. and Larson, Allan and Losos, Jonathan B.},
  title   = {Tempo and mode of evolutionary radiation in iguanian lizards},
  journal = {Science},
  year    = {2003},
  volume  = {301},
  number  = {5635},
  pages   = {961--964},
  doi     = {10.1126/science.1084786}
}

@article{Imhof1961,
    author = {Imhof, J. P.},
    title = {Computing the distribution of quadratic forms in normal variables},
    journal = {Biometrika},
    volume = {48},
    number = {3-4},
    pages = {419-426},
    year = {1961},
    month = {12},
    issn = {0006-3444},
    doi = {10.1093/biomet/48.3-4.419},
    url = {https://doi.org/10.1093/biomet/48.3-4.419}
}

@article{Kendall1948a,
  title={On the generalized ``birth-and-death'' process},
  author={Kendall, D.G.},
  journal={The annals of mathematical statistics},
  volume={19},
  number={1},
  pages={1--15},
  year={1948},
  publisher={Institute of Mathematical Statistics}
}

@article{Kendall1948b,
  title={On some modes of population growth leading to {RA Fisher}'s logarithmic series distribution},
  author={Kendall, D.G.},
  journal={Biometrika},
  volume={35},
  number={1/2},
  pages={6--15},
  year={1948},
  publisher={JSTOR}
}

@book{Lewin1982,
      author        = "Lewin, Leonard",
      title         = "Polylogarithms and associated functions",
      publisher     = "North-Holland",
      address       = "New York",
      year          = "1981"
}

@book{Mathai1992,
author = {Mathai, A. M. and Provost, Serge B.},
address = {New York},
booktitle = {Quadratic forms in random variables : theory and applications},
isbn = {0824786912},
language = {eng},
lccn = {91040165},
publisher = {Dekker},
series = {Statistics, textbooks and monographs ; v.126},
title = {Quadratic forms in random variables : theory and applications},
year = {1992},
}

@article {Morris1979,
    AUTHOR = {Morris, Robert},
     TITLE = {The dilogarithm function of a real argument},
   JOURNAL = {Math. Comp.},
  FJOURNAL = {Mathematics of Computation},
    VOLUME = {33},
      YEAR = {1979},
    NUMBER = {146},
     PAGES = {778--787},
      ISSN = {0025-5718,1088-6842},
   MRCLASS = {65D20 (33A70)},
  MRNUMBER = {521291},
       DOI = {10.2307/2006312},
       URL = {https://doi.org/10.2307/2006312},
}

@article{Mulder2015,
title = {On the distribution of interspecies correlation for {M}arkov models of character evolution on {Y}ule trees},
journal = {Journal of Theoretical Biology},
volume = {364},
pages = {275-283},
year = {2015},
issn = {0022-5193},
doi = {https://doi.org/10.1016/j.jtbi.2014.09.016},
url = {https://www.sciencedirect.com/science/article/pii/S0022519314005554},
author = {Willem H. Mulder and Forrest W. Crawford}
}

@article{Nee1994,
  title={Extinction rates can be estimated from molecular phylogenies},
  author={Nee, S. and Holmes, E.C. and May, R.M. and Harvey, P.H.},
  journal={Philosophical Transactions of the Royal Society of London. Series B: Biological Sciences},
  volume={344},
  number={1307},
  pages={77--82},
  year={1994},
  publisher={The Royal Society}
}

@article{OMeara2006,
  author  = {O'Meara, Brian C. and An{\'e}, C{\'e}cile and Sanderson, Michael J. and Wainwright, Peter C.},
  title   = {Testing for different rates of continuous trait evolution using likelihood},
  journal = {Evolution},
  year    = {2006},
  volume  = {60},
  number  = {5},
  pages   = {922--933},
  doi     = {10.1111/j.0014-3820.2006.tb01171.x}
}

@article{Sagitov2012,
title = {Interspecies correlation for neutrally evolving traits},
journal = {Journal of Theoretical Biology},
volume = {309},
pages = {11-19},
year = {2012},
issn = {0022-5193},
doi = {https://doi.org/10.1016/j.jtbi.2012.06.008},
url = {https://www.sciencedirect.com/science/article/pii/S0022519312002858},
author = {Serik Sagitov and Krzysztof Bartoszek}
}

@Article{Sagitov2013,
author={Sagitov, Serik
and Shaimerdenova, Altynay},
title={Extinction times for a birth--death process with weak competition},
journal={Lithuanian Mathematical Journal},
year={2013},
month={Apr},
day={01},
volume={53},
number={2},
pages={220-234},
issn={1573-8825},
doi={10.1007/s10986-013-9204-x},
url={https://doi.org/10.1007/s10986-013-9204-x}
}

@article{Stadler2009,
title = {On incomplete sampling under birth-death models and connections to the sampling-based coalescent},
journal = {Journal of Theoretical Biology},
volume = {261},
number = {1},
pages = {58-66},
year = {2009},
issn = {0022-5193},
doi = {https://doi.org/10.1016/j.jtbi.2009.07.018},
url = {https://www.sciencedirect.com/science/article/pii/S0022519309003300},
author = {Tanja Stadler}
}

@article{Tavare2025,
    author = {Tavaré, Simon},
    title = {Birth and death processes in phylogenetics and population genetics},
    journal = {Philosophical Transactions of the Royal Society B: Biological Sciences},
    volume = {380},
    number = {1919},
    pages = {20230300},
    year = {2025},
    month = {02},
    issn = {0962-8436},
    doi = {10.1098/rstb.2023.0300},
    url = {https://doi.org/10.1098/rstb.2023.0300}
}

@article{Wills1994, title={Disparity as an evolutionary index: a comparison of Cambrian and Recent arthropods}, volume={20}, DOI={10.1017/S009483730001263X}, number={2}, journal={Paleobiology}, author={Wills, Matthew A. and Briggs, Derek E. G. and Fortey, Richard A.}, year={1994}, pages={93–130}}

@incollection{Zagier2007,
author="Zagier, Don",
editor="Cartier, Pierre
and Moussa, Pierre
and Julia, Bernard
and Vanhove, Pierre",
title="The {D}ilogarithm {F}unction",
bookTitle="Frontiers in Number Theory, Physics, and Geometry II: On Conformal Field Theories, Discrete Groups and Renormalization",
year="2007",
publisher="Springer Berlin Heidelberg",
address="Berlin, Heidelberg",
pages="3--65",
isbn="978-3-540-30308-4",
doi="10.1007/978-3-540-30308-4_1",
url="https://doi.org/10.1007/978-3-540-30308-4_1"
}



\end{document}